\PassOptionsToPackage{verbose=true,margin=1in,letterpaper}{geometry}

\documentclass{article}
\usepackage{geometry}

\usepackage{PRIMEarxiv}

\usepackage[utf8]{inputenc} % allow utf-8 input
\usepackage[T1]{fontenc}    % use 8-bit T1 fonts
\PassOptionsToPackage{hyphens}{url}
\usepackage{hyperref}       % hyperlinks
\usepackage{booktabs}       % professional-quality tables
\usepackage{amsfonts}       % blackboard math symbols
\usepackage{nicefrac}       % compact symbols for 1/2, etc.
\usepackage{microtype}      % microtypography
\usepackage{lipsum}
\usepackage{fancyhdr}       % header
\usepackage{graphicx}       % graphics
\usepackage{subcaption}
\graphicspath{{media/}}     % organize your images and other figures under media/ folder
\usepackage{multirow}
\usepackage{makecell}
\usepackage[margin=1in]{geometry}
\usepackage{array}
\usepackage{longtable}
\usepackage{caption}
\usepackage{tabularx}
\usepackage{booktabs}
\usepackage{ragged2e}

\usepackage{enumitem}

\usepackage{verbatim}       % for comment
\usepackage{xcolor}         % for textcolor
\usepackage{amsmath}

\usepackage{float}

\usepackage{multirow}

\usepackage{soul}

\usepackage{subcaption}     % for subfigure

\usepackage{longtable}      % for long table

\usepackage{colortbl}       % for row color

\newcommand{\liming}[1]{\textcolor{teal}{\textbf{Liming:} #1}}
\newcommand{\sunny}[1]{\textcolor{blue}{\textbf{Sunny:} #1}}

\title{Software Security Mapping Framework: Operationalization of Security Requirements}

\author{
  Sung Une Lee, Liming Dong, Zhenchang Xing, Muhammad Ejaz Ahmed\\
  Data61, CSIRO, Australia\\
     \AND
  Stefan Avgoustakis \\
  Google, Australia \\
}

\begin{document}
\maketitle

\begin{abstract}
The escalating complexity of modern software development environments has heightened concerns around supply chain security. Yet, existing frameworks often fall short in translating abstract security principles into concrete, actionable practices. This paper introduces the Software Security Mapping Framework, a structured solution designed to operationalize security requirements across hierarchical levels—from high-level regulatory standards (e.g., ISM, Australia's official cybersecurity standard published by the Australian Signals Directorate (ASD)), through mid-level frameworks (e.g., NIST SSDF, the US secure software development framework), to fine-grained technical activities (e.g., SLSA, software supply chain security frmework).
Developed through collaborative research with academic experts and industry practitioners, the framework systematically maps 131 refined security requirements to over 400 actionable operational steps spanning the software development life-cycle. It is grounded in four core security goals: Secure Software Environment, Secure Software Development, Software Traceability, and Vulnerability Management. Our approach leverages the KAOS goal modeling methodology to establish traceable linkages between strategic goals and tactical operations, enhancing clarity, accountability, and practical implementation. 
To facilitate adoption, we provide a web-based navigation tool for interactive exploration of the framework. A real-world case study based on the Log4j vulnerability illustrates the framework’s practical utility by generating a tailored checklist aligned with industry best practices. In addition, we offer a structured, machine-readable OSCAL Catalog Model of the Software Security Mapping Framework, enabling organizations to automate implementation, streamline compliance processes, and effectively respond to evolving security risks.

\end{abstract}

% keywords can be removed
\keywords{Software supply-chain \and security \and }

\section{Introduction} \label{sec:introduction}

The security of the software supply chain has emerged as a critical concern in today's interconnected and rapidly evolving digital landscape. The complexity of the software supply chain, combined with the growing number of stakeholders involved in the software ecosystem, has significantly increased the risk of vulnerabilities and attacks. A seemingly minor compromise at any stage of the chain can result in the complete subversion of the final product, underscoring the necessity for robust security measures \cite{Zahan2023Software, Torres2020intoto}. In response to these challenges, organizations often turn to official security standards, frameworks, and regulations to guide and enhance their security practices \cite{kalu2024industry}. However, despite the existence of these resources, national regulations remain insufficient to enforce compliance with secure software supply chain practices, leaving significant gaps in global security efforts \cite{Kaspar2022Preventing}.
    
Developing practical and effective solutions for software supply chain security from a holistic perspective presents notable challenges \cite{melara2022what}. Existing security regulations/guidelines/frameworks, while valuable, are often criticized for being overly generic and failing to address the specific needs of software development teams and engineers. As practitioners have noted, many guidelines lack actionable details and fail to provide concrete, universally applicable rules \cite{Sammak2024Developers}. Moreover, the integration of security standards and regulatory requirements into the software development life-cycle (SDLC) proves difficult in practice. Challenges such as insufficient training for engineers and technical limitations in implementing these standards further hinder efforts to achieve secure software supply chain practices \cite{kalu2024industry}.
    
These challenges highlight the need for systematic solutions that incorporate a holistic mapping of software security frameworks to detailed operational security requirements to enhance the security of the software supply chain. Such solutions are essential for promoting the practical implementation of software security frameworks in real-world SDLC, effectively addressing the complexity of software supply chain security risks and overcoming the practical obstacles faced by organizations and individual practitioners.
    
The security of the software supply chain has garnered increasing attention in recent years, driven by its critical role in modern software development and the growing prevalence of supply chain attacks. Researchers have explored various frameworks and methodologies to address these challenges. Sun et al.\cite{SunQYZM24} proposed a knowledge-driven framework that systematically analyzes software supply chain security risks, emphasizing structured approaches to identify and mitigate vulnerabilities. Hassanshahi et al.\cite{HassanshahiMMSB23} introduced a logic-based framework for ensuring supply chain security assurance through dependency modeling and trust evaluation. 
Several studies\cite{melara2022what,Sammak2024Developers,EnckW22} have examined the practical challenges faced by developers and organizations in implementing secure supply chain practices. For example, Sammak et al.\cite{Sammak2024Developers} conducted an interview-based study, uncovering gaps between existing security guidelines and the specific needs of developers, highlighting the limitations of overly generalized software security regulations/guidelines/frameworks. 
    
The goal of this work is to assist companies and individual developers in effectively navigating and implementing software supply chain security requirements by addressing the fragmented nature of existing regulations, frameworks, and practices. Through a collaborative research study, we aim to provide a holistic mapping that spans three levels frameworks until to detailed operational steps—offering practical reference architecture model that is easy to follow and implement in diverse real-world scenarios. This study also provides a machine-readable format for this mapping, which enables better portability, integration with automation tools, and support for continuous compliance and assessment workflows.

This study makes key contributions by i) providing a multi-layered and goal-driven mapping framework that connects high-level software security goals to operational actions through structured decomposition, ii) enabling operationalization and interoperability by deriving over 400 detailed operations and representing them in an extended, machine-readable data format, enabling traceability, automation, and tool integration, iii) offering comprehensive coverage through the integration of key existing frameworks and a real-world application, demonstrating practical utility across the entire software supply chain, and iv) laying the groundwork for future research in security requirements traceability and life-cycle-aware security modeling.

The remainder of this paper is organized as follows. 
Section \ref{sec:background} introduces background knowledge about software supply chain security and existing frameworks and mappings.
Section \ref{sec:Methodology} provides details of the methodoloy used to conduct the software security mapping. 
Section \ref{sec:Mapping} presents structure of holistic mapping that spans three levels, from high-level frameworks to detailed operational steps to provide supply chain security mapping framework. 
Section \ref{sec:Discussion} presents the practical applications of the mapping framework, introduces the machine-readable data format, and discusses the theoretical implications of this study.
Section \ref{sec:Conclusion} concludes this study and discusses potential directions for future work.

\section{Background} \label{sec:background}
%\liming{I've removed the 2.4 and kept mapping definition at 2.3}

Before delving into specific frameworks and practices, it is important to understand the broader context in which software supply chain security has emerged as a critical concern. This section outlines the core security risks that arise from the complexity of modern software supply chains and reviews key frameworks and existing mapping efforts that have been proposed to mitigate security risks. Together, these insights provide the foundation for our multi-layered and goal-driven mapping framework, which aims to bridge the gap between high-level governance requirements and low-level technical implementation.

\subsection{Software Supply Chain Security Risks} \label{subsec:Security Risks}

Modern software supply chains consist of complex and interconnected processes, components, and stakeholders. This complexity introduces significant challenges to ensuring the delivery of secure software products and services. As a result, the software supply chain has become a critically vulnerable attack surface, experiencing a significant surge in software security risks~\cite{Donoghue2024SBOM}.

Key risks include insecure development environments, unpatched or malicious dependencies, and a lack of visibility into component provenance. Poorly secured environments can lead to unauthorized access across development, testing, and production stages~\cite{BauerCRRV09}, while reliance on outdated or unverified packages propagates vulnerabilities across the supply chain~\cite{YanNLLLB21,ZahanZGMMW22,Wang0HSX0WL20}. Attackers may also inject malicious code via compromised tools or third-party libraries~\cite{LiWFWWW23}.

The lack of component transparency, particularly regarding transitive dependencies, has driven the adoption of practices like Software Bills of Materials (SBOMs)~\cite{Donoghue2024SBOM}. Vulnerability management remains one of the most challenging areas, as organizations must prioritize which of the many reported vulnerabilities require urgent action~\cite{Yin2023,Smyth17}.

Historically, software supply chains were not seen as deliberate attack vectors \cite{williams2024proactive}, while the increasing complexity and interdependence of components have fundamentally changed the threat landscape. Adversaries now actively exploit this complexity by implanting vulnerabilities into upstream dependencies or compromising development and deployment infrastructure. These attacks exploit the very interconnectedness that makes modern software delivery efficient, allowing malicious code to spread rapidly across systems and organizations. As awareness of these systemic risks grows, governments and industry stakeholders respond with frameworks and practices that aim to increase transparency, strengthen provenance, and reduce the likelihood and impact of supply chain compromises.

\subsection{Existing Software Security Frameworks} \label{subsec: Existing Frameworks}

In response to the growing threats to the software supply chain, a range of standards, frameworks, and guidelines have emerged, each addressing different facets of software security. Notable examples include ISM, NIST SSDF, SLSA, TUF, SAMM, and S2C2F. While each offers valuable insights and controls, they vary significantly in scope, depth, and applicability—creating a fragmented landscape that poses challenges for consistent interpretation and adoption across organizations.

\textbf{ISM (Information Security Manual) \footnote{ISM: Guidelines for Software Development. \url{https://www.cyber.gov.au/resources-business-and-government/essential-cyber-security/ism/cyber-security-guidelines/guidelines-software-development}}:}  
Developed by the Australian Signals Directorate, ISM provides high-level security guidelines, including controls relevant to software development. While valuable for aligning with governance and compliance, ISM is abstract in nature and lacks concrete implementation guidance, making it difficult to operationalize within typical development life-cycles.

\textbf{NIST SSDF (Secure Software Development Framework) \footnote{NIST: Secure Software Development Framework (SSDF) Version 1.1. \url{https://csrc.nist.gov/pubs/sp/800/218/final}}:}  
NIST SSDF organizes secure software development practices into four groups: Prepare the Organization, Protect the Software, Produce Well-Secured Software, and Respond to Vulnerabilities. While the framework provides clear security tasks across the SDLC. As a framework-agnostic baseline, SSDF is applicable to organizations of all sizes and sectors, regardless of their development methodology or toolchain. However, while SSDF specifies what security practices should be in place, it does not prescribe how to implement them—leaving implementation details to be determined by the adopting organization.

\textbf{SLSA (Supply-chain Levels for Software Artifacts) \footnote{SLSA V1.0. \url{https://slsa.dev/spec/v1.0/}}:}  
SLSA defines a tiered model to secure software artifacts, focusing on verifiable build provenance and tamper resistance. It introduces specific technical requirements related to build systems and CI/CD pipelines. While detailed and actionable, its focus is limited to post-development build integrity and lacks broader guidance on software governance and early-stage development practices.

\textbf{TUF (The Update Framework) \footnote{TUF. \url{https://theupdateframework.io}}:}  
TUF aims to secure the software update process, even in cases where signing keys or repositories are compromised. It provides robust cryptographic and metadata-based protections. However, its scope is narrowly focused on update infrastructure and does not address broader supply chain concerns such as development processes, dependency management, or life-cycle governance.

\textbf{SAMM (Software Assurance Maturity Model) \footnote{SAMM. \url{https://owaspsamm.org}}:}  
OWASP SAMM is a maturity model that supports the evaluation and continuous improvement of an organization’s software security posture. It is structured around five business functions and 15 security practices, each organized across three maturity levels acticities. SAMM could serves as a practical framework for operationalizing high-level standards such as NIST SSDF.

\textbf{S2C2F (Secure Supply Chain Consumption Framework) \footnote{S2C2F. \url{https://github.com/ossf/s2c2f/tree/main}}:}  
S2C2F focuses on securing the consumption of open-source software (OSS) packages. It outlines eight practices across four maturity levels, based on known adversary techniques. While its scope is limited to OSS consumption rather than the full development life-cycle, S2C2F offers actionable, threat-informed guidance that helps organizations strengthen their dependency management and reduce risks associated with third-party software.

While each of the aforementioned frameworks offers valuable guidance on particular aspects of software supply chain security, they often address different layers of the ecosystem in a fragmented landscape that can be challenging to navigate in practice. 
ISM provides governance-level controls but lacks implementation specificity. NIST SSDF guidance aim to cover SDLC practices, yet still leave gaps in translating high-level requirements into operational processes. SLSA and TUF provide technical depth in narrow domains, such as build integrity and update security, but do not address broader SDLC or policy concerns.

This fragmentation creates a barrier for organizations attempting to apply software supply chain security holistically. It remains difficult to trace how a high-level policy (e.g., an ISM control) relates to concrete actions, such as provenance requirements in SLSA. A unified mapping is needed to connect governance-level intent with technical and operational execution.

\subsection{Software Security Mapping} \label{subsec: Existing Mapping}

In this study, we define "Software Security Mapping" as the process of identifying, aligning, and integrating key concepts, objectives, control, and recommended practices from existing software supply chain security frameworks into a unified, structured reference architecture.

To support the adoption of secure software development guidance, several mapping efforts have been introduced to align overlapping frameworks and identify conceptual equivalences. While these efforts provide valuable reference points, they typically focus on specific pairwise comparisons and lack deeper integration across diverse security frameworks. Moreover, they often fall short of offering a unifying structure that spans the full software development life-cycle. Below, we summarize the most relevant examples.

\textbf{SSDF mapping to SAMM \footnote{NIST SSDF mapping to SAMM. \url{https://owaspsamm.org/blog/2023/02/06/samm-ssdf-mapping/}}:}  
The OWASP SAMM team has developed a bidirectional mapping between NIST SSDF tasks and SAMM activities. This linkage helps connect SSDF's broad security practices to SAMM’s structured activities and maturity levels, providing a pathway for organizations to operationalize SSDF through SAMM. While conceptually sound, the mapping lacks granularity for integration into tooling or operational workflows.

\textbf{SLSA mapping to SSDF \footnote{SLSA mapping to Other Frameworks. \url{https://slsa.dev/blog/2022/07/slsa-foundational-framework}}:}  
The SLSA working group aligned its build-level requirements with related SSDF tasks. This mapping offers technical depth in securing build pipelines and artifact provenance but is limited in life-cycle coverage and lacks broader governance alignment.

\textbf{S2C2F mapping to Other Frameworks \footnote{S2C2F mapping to Other Frameworks. \url{https://github.com/ossf/s2c2f/blob/main/specification/framework.md}}:}  
S2C2F maps its OSS-focused practices to multiple specifications, including SSDF, SLSA, and CIS. While this supports OSS governance, the mapping is narrow in scope and does not address broader development and deployment concerns.

\textbf{P-SSCRM mapping to Ten Industry Frameworks:}  
The Proactive Software Supply Chain Risk Management Framework (P-SSCRM) Version 1.0~\cite{williams2024proactive} outlines 73 risk management tasks across 15 practices, grouped by product life-cycle stages: Governance, Product, Environment, and Deployment. It maps these tasks to ten existing standards, SSDF, SLSA, BSIMM, OpenSSF, and OWASP SCVS etc. Although broad in scope, the P-SSCRM serves more as a strategic reference. It lacks deeper operational guidance and fine-grained integration between frameworks, making it less suitable for direct implementation.

\section{Methodology} \label{sec:Methodology}

\begin{comment}
\begin{itemize}
    \item The main methodology is "collaborative research"; explain the key roles (e.g., practitioner (Google), security experts (Seyit and Ejaz), software engineers/researchers (Zhenchang, Liming, Sunny), and so on).
    \item Include Goal Oriented Requirement Engineering (GORE) framework- specifically, KAOS approach.
    \item Mention mapping.
    \item Include literature review (?); need to decide this in Jan.
    \item Include evaluation results (internal, Google team, and external reviewers?); may be in Jan.
\end{itemize}
\end{comment}

%\sunny{The following is about "collaborative research"}

%\sunny{Include 1-2 sentences that mention ISM is an example regulation used in this study, but not limited to this; any regulations can be used.... Consider this statement in the introduction/conclusion, not here.}

We adopted a collaborative research methodology~\cite{ward1982collaborative} to develop a software security framework, emphasizing ongoing dialogue and shared objectives. This study involved active participation from both researchers and industry experts in the fields of software engineering and security, guided by three essential principles of collaborative research. 
First, researchers and practitioners collaborated closely throughout the process. 
Second, the approach prioritized both practical problem-solving and theoretical advancements. 
Finally, the participants cultivated mutual respect and enhanced their knowledge and understanding of software security practices and operations.

%\sunny{The following is about formation of the research team.}

The research team consisted of seven core members: five researchers and two industry experts. Additional contributors participated in specific tasks, such as conducting reviews and providing specialized input when needed.
The industry experts work for a global-leading technology company. With extensive experience in software engineering and security, they contributed valuable insights into industry practices and helped align the research with real-world challenges and practical solutions.

Table \ref{tab:researchteam} outlines the roles and responsibilities assigned to each core team member for this study.

\begin{table} [htb]
 \footnotesize
    \begin{tabular}{p{0.15\textwidth}p{0.31\textwidth}p{0.46\textwidth}}
    \hline
   Role & Position & Responsibility\\ 
    \hline
         Project lead & Senior researcher in software and security & Oversee the project direction and progress. \\
         Project manager & Senior researcher in software and security & Manage/lead project activities, including defining scope, delivering outputs, and reviewing research results. \\
         Research lead  & Senior researcher in software and security & Contribute to designing the research methodology and reviewing findings. \\
         Researcher (1) & Researcher in software and security & Conduct designing the research methodology, data collection, and reviewing literature.\\
         Researcher (2) & Researcher in software and security & Conduct data collection, reviewing literature, and supporting analysis and documentation. \\
         Industry expert (1) & Security practice lead & Share industrial issues and problems in software security and provide industry requirements and potential solutions. \\
         Industry expert (2) & Software and security engineer & Provide industry-specific expertise and practical insights, bridging perspectives from researchers and practitioners. \\
    \hline
    \end{tabular}

    \caption{Research team - roles and responsibilities}
    \label{tab:researchteam}
\end{table}

%\sunny{Include a overall process (diagram) and emphasize the practitioners regularly engaged; had bi-weekly meetings; had reviews.....}

\begin{figure*} [htb]
    \centering
    \includegraphics[width=\textwidth]{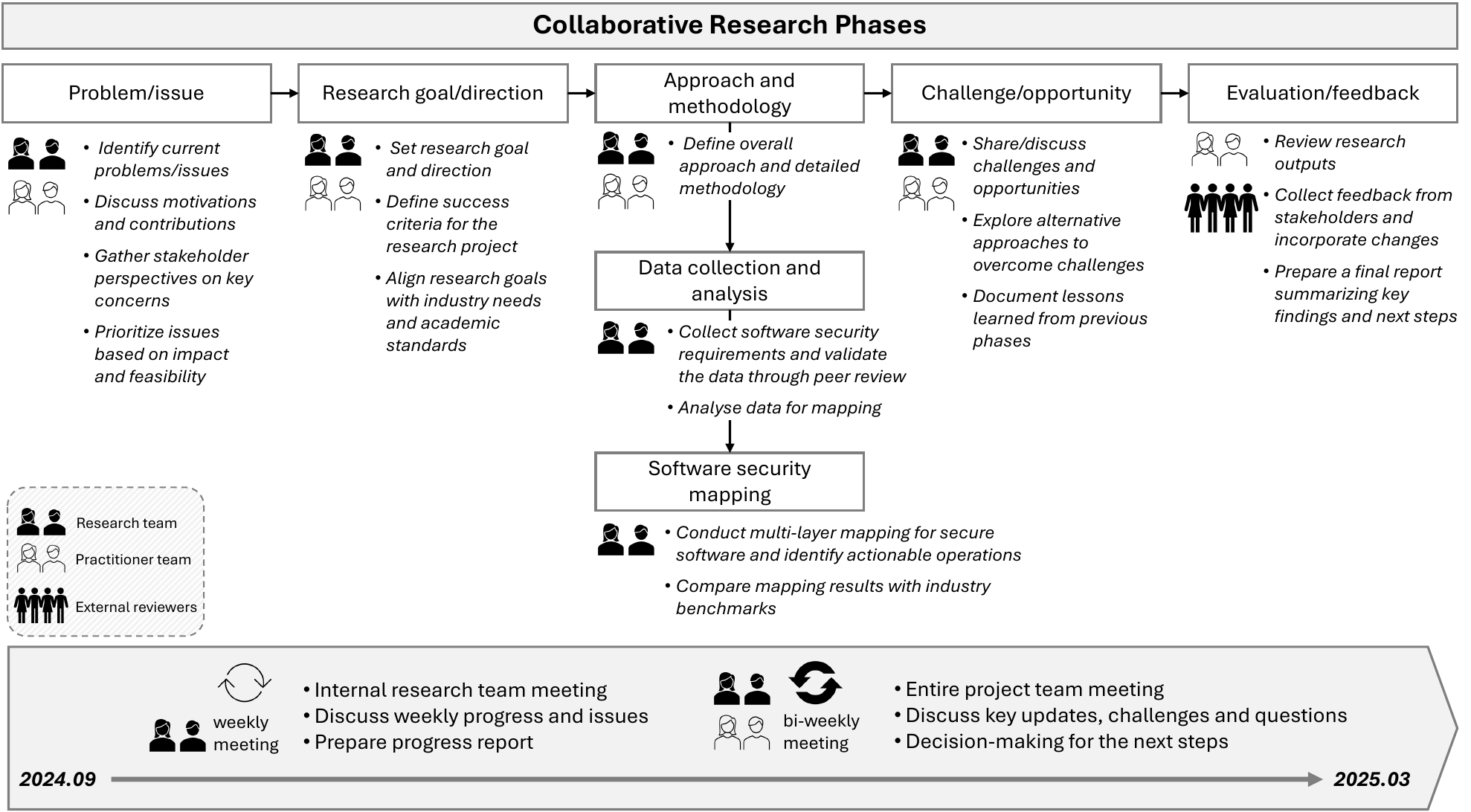}
    \caption{The collaborative research phases implemented in this study.}
    \label{fig:collaborative research}
\end{figure*}

The collaborative research process followed in this study is depicted in Figure \ref{fig:collaborative research}, which outlines seven key phases: Problem/Issue, Research Goal/Direction, Approach and Methodology, Challenge/Opportunity, Data Collection and Analysis, Software Security Mapping, and Evaluation/Feedback.

\textbf{Problem/Issue:}
The research team began by identifying current problems and issues in software security practices. This phase also involved discussions on motivations and contributions from stakeholders, gathering diverse perspectives on key concerns, and prioritizing issues based on their potential impact and feasibility.

\textbf{Research Goal/Direction:}
Once the issues were identified, we collaboratively set the research goals and direction. Success criteria for the research project were defined, ensuring alignment with both industry needs and academic standards.
The success criteria of this study includes practical applicability, collaborative engagement, stakeholder satisfaction, and knowledge dissemination.

\textit{Practical applicability} means that the research outcomes should be directly applicable in real-world settings.
Industry partners should be able to implement the developed mapping framework in their software security practices. For this, we aimed to identify actionable operations through the software security mapping framework. 

\textit{Collaborative engagement} was a focal success factor for this study. We implemented sustained and meaningful collaboration between researchers and industry practitioners throughout the project.
Additionally, we had regular feedback loops and iterative improvements based on stakeholder input.

\textit{Stakeholder satisfaction} is to ensure that the key stakeholders (both researchers and industry experts) find value in the research outcomes.
In this study, we addressed their key concerns and challenges, and obtained positive feedback from stakeholders through a final project evaluation survey %\sunny{this should be reviewed/revised after evaluation}.

\textit{Knowledge dissemination} refers that the research findings/outputs should be shared with the broader research and practitioner communities.
This can be generally achieved through presentations, workshops, or open access publications. We plan to provide open access web-pages to share the research outputs %\sunny{this should be reviewed/revised after evaluation}.

\textbf{Approach and Methodology:}
We established the overall research approach and designed a detailed methodology to conduct this study. Peer review was integrated to validate the methodology.

%\sunny{About GORE/KAOS}
In this study, we focused on identifying software security requirements and actionable operations based on the requirements.
To do so, we adopted goal oriented requirements engineering (GORE).
GORE has been used to effectively identify requirements and achieve goals \cite{van2001goal}. 
We specifically used KAOS approach to identify goals, requirements, and operations for practitioners.
This approach is a formal method for modeling and reasoning about system requirements based on goals \cite{werneck2009comparing}. 
It helps break down high-level strategic goals into finer-grained requirements through goal decomposition. There are several key components in KAOS:
\begin{itemize}
    \item Goals: High-level objectives or purposes (what the system should achieve).
    \item Requirements: Specific conditions or capabilities needed to meet those goals.
    \item Operationalizations: Technical and practical details that define how the system will achieve the requirements.
    \item Agents: Stakeholders or systems responsible for fulfilling goals and requirements.
\end{itemize}

\begin{figure*}
    \centering
    \includegraphics[width=0.8\textwidth]{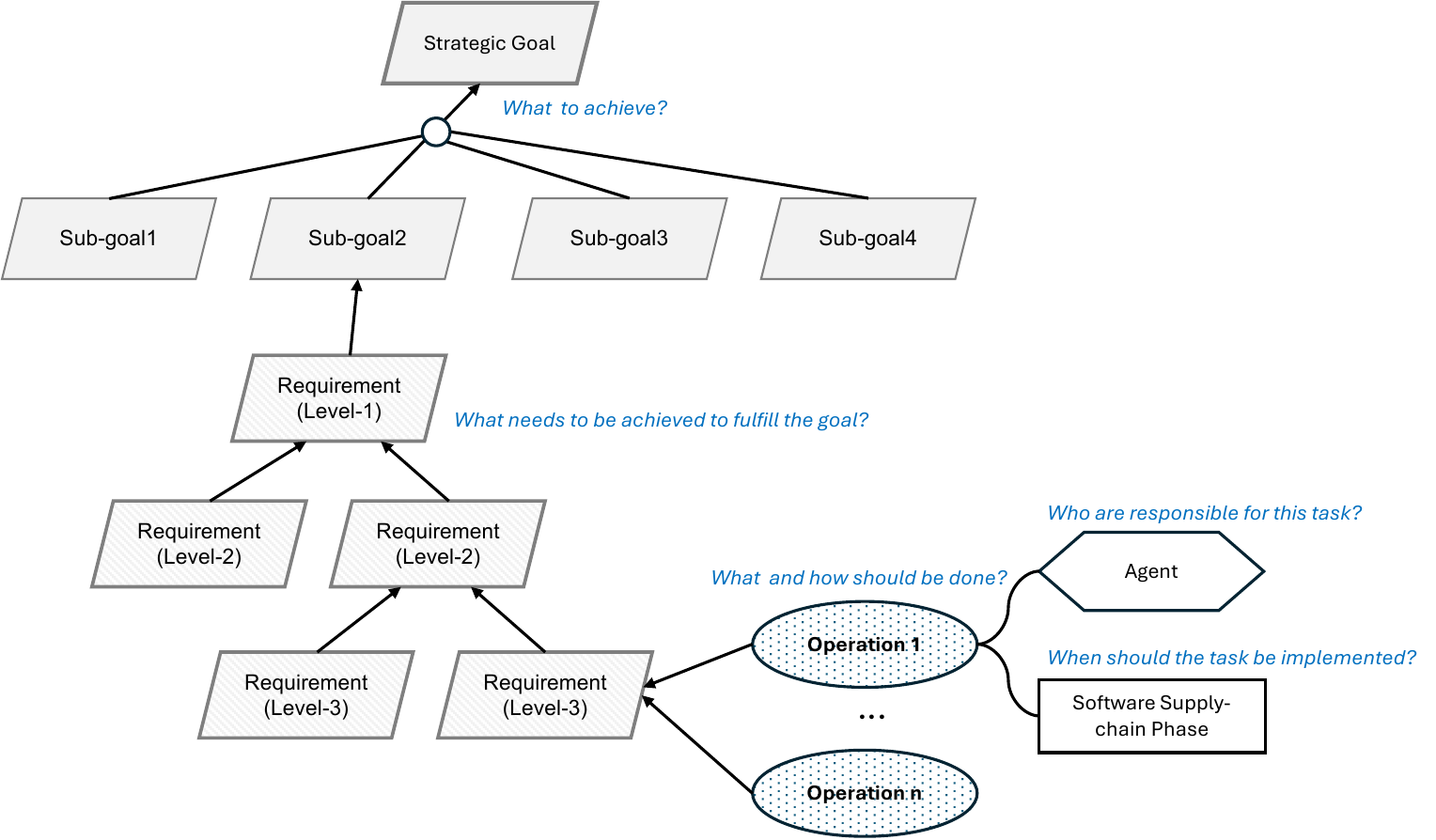}
    \caption{Overview of the approach for operationalizing software security requirements in this study.}
    \label{fig:operation}
\end{figure*}

Figure \ref{fig:operation} shows an overview of the approach we used in this study. 
The figure illustrates the hierarchical breakdown from strategic goals to sub-goals, requirements at different levels, and operations, specifying agents responsible for tasks and the phases of the software supply chain during which implementation should occur.

\textbf{Challenge/Opportunity:}
The Challenge/Opportunity phase focused on identifying and discussing key challenges and potential opportunities related to software security. The research team explored alternative approaches to address the identified challenges and documented lessons learned from earlier phases to refine and enhance the research framework. This phase was conducted bi-weekly with the participation of the entire team to ensure continuous feedback and alignment.

The primary challenges encountered during this study were:

\textit{Lack of completeness in the primary frameworks:}

Initially, we used ISM, NIST SSDF, SLSA, and TUF as the primary frameworks. However, the first round of mapping revealed significant gaps or missing requirements, represented by empty cells in the traceability matrix.

\textit{High volume of requirements and operations:}

The number of identified requirements and operations increased rapidly, raising concerns about feasibility and manageability.

To address the gaps in the initial mapping, we conducted a broader survey of additional frameworks and resources. These included CISA, SAMM, NIST AI RMF, NIST GenAI, and OSSF S2C2F. By incorporating these additional frameworks, we were able to fill the missing requirements and practices. After four iterative rounds of mapping, we achieved a 100\% completion rate, a significant improvement from the initial 13\% completion rate in the first round.

While this iterative mapping process helped achieve comprehensive coverage, we identified a total of 131 Level-3 requirements. With an estimated 4 to 5 operations per requirement, the total number of operations was expected to exceed 600. A feasibility test conducted on several requirements revealed potential complexity issues, making it challenging to manage such a high volume of operations effectively.

To enhance understanding, cohesion, and manageability, we proposed narrowing our focus by applying specific criteria to refine and prioritize the key requirements:

\begin{itemize}
    \item Relevance: Requirements that were not directly aligned with the strategic goal or higher-level requirements were excluded.
    \item Overlap: Requirements that overlapped or had similar concepts within the same requirement group were consolidated or excluded to avoid redundancy.
    \item Feasibility: Requirements that were too high-level, broad, or difficult to operationalize were excluded to focus on actionable and practical requirements.
\end{itemize}

By applying these criteria, we reduced the overall volume of requirements while increasing the coherence and manageability of the final mapping. This approach ensures that the remaining requirements are both relevant and feasible for practical implementation.

\textbf{Data Collection and Analysis:}
We collected software security requirements from existing frameworks and incorporated practical insights through direct engagement with industry practitioners. The collected data was validated through peer review, ensuring accuracy and relevance. A detailed analysis was then conducted to support the development of the software security mapping.

Table \ref{tab:frameworks} presents the key frameworks selected for data collection and analysis, the total number of collected data, and their primary usage in this study. The collected data from each framework has been analyized in depth, with some further broken down to provide fine-grained requirements and operations, while others have been merged to ensure consistency and eliminate redundancies.

\begin{table} [htb]
    \centering
    \footnotesize
    \begin{tabular}{p{0.1\textwidth}p{0.45\textwidth}p{0.2\textwidth}p{0.15\textwidth}}
    \hline
    Framework & Description & Data collected & Used for \\ 
    \hline
    ISM & Information Security Manual; An Australia cyber security framework that an organisation can apply, to protect their information technology and operational technology systems, applications and data from cyber threats. & 23 regulatory requirements for software development  & Goal, Level-1 \\
    SSDF & NIST Secure Software Development Framework; A set of fundamental, sound, and secure software development practices based on established secure software development practice documents from organizations. & 42 software development practices & Level-2\\
    SLSA & Supply-chain Levels for Software Artifacts; A security framework, a checklist of standards and controls to prevent tampering, improve integrity, and secure packages and infrastructure. & 7 level 1,2,3 checklists & Level-3, Operation \\
    SAMM & The Software Assurance Maturity Model; An open framework to help organizations formulate and implement a strategy for software security. & 90 activities from 15 practices & Level-3, Operation \\
    TUF & The Update Framework; It maintains the security of software update systems, providing protection even against attackers that compromise the repository or signing keys. & 27 relevant specifications & Level-3, Operation \\
    S2C2F & Secure Supply Chain Consumption Framework; It is designed based on known threats (i.e. tactics and techniques) used by adversaries to compromise Open Source Software (OSS) packages. & 25 requirements from 8 practices & Level-3, Operation \\
    \hline
    \end{tabular}
    \caption{The existing software security frameworks primimarily used in this study; There are some complementary framework used such as CISA Securing the Software Supply Chain, and OWASP famework, NIST AI Risk Management Framework.}
    \label{tab:frameworks}
    \normalfont
\end{table}

%We identified 23 security requirements for software development from Australian Information Security Manual (ISM) for this level.

%NIST Secure Software Development Framework (SSDF) is an example framework of this level. 

%The Supply-chain Levels for Software Artifacts (SLSA) and The Update Framework (TUF) are the example frameworks for this level. 

\textbf{Software Security Mapping:}
Based on the collected data and analysis, we developed a multi-level framework for secure software. This mapping was conducted using the seven steps previously introduced, from \textit{"Define Top-Most Strategic Goal"} to \textit{"Identify Operations and Agents"}.

%\sunny{include a overview of the mapping, such as statistical information (graph, table, etc.); need to differentiate the next section, overview.}

\begin{figure} [htb]
    \centering
    \begin{subfigure}{0.9\textwidth}
    \includegraphics[width=\textwidth]{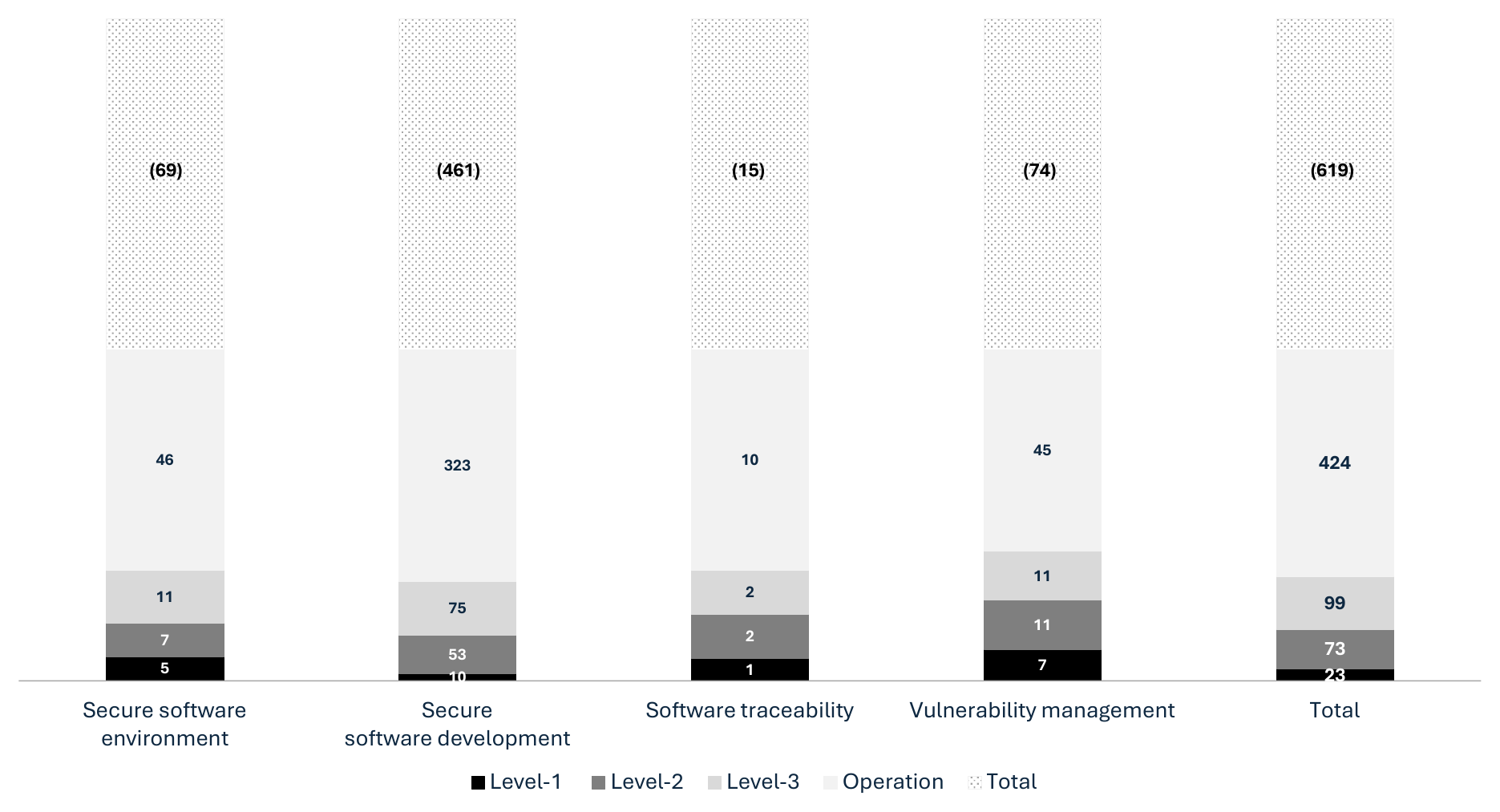}
    \caption{The distribution of requirements and operations across software security goals, illustrating the breakdown by levels (Level-1, Level-2, Level-3) and operations under each goal. The total count of requirements and operations for each goal is also highlighted.}
    \label{fig:statistic1}        
    \end{subfigure}
    \begin{subfigure}{0.9\textwidth}
    \includegraphics[width=\textwidth]{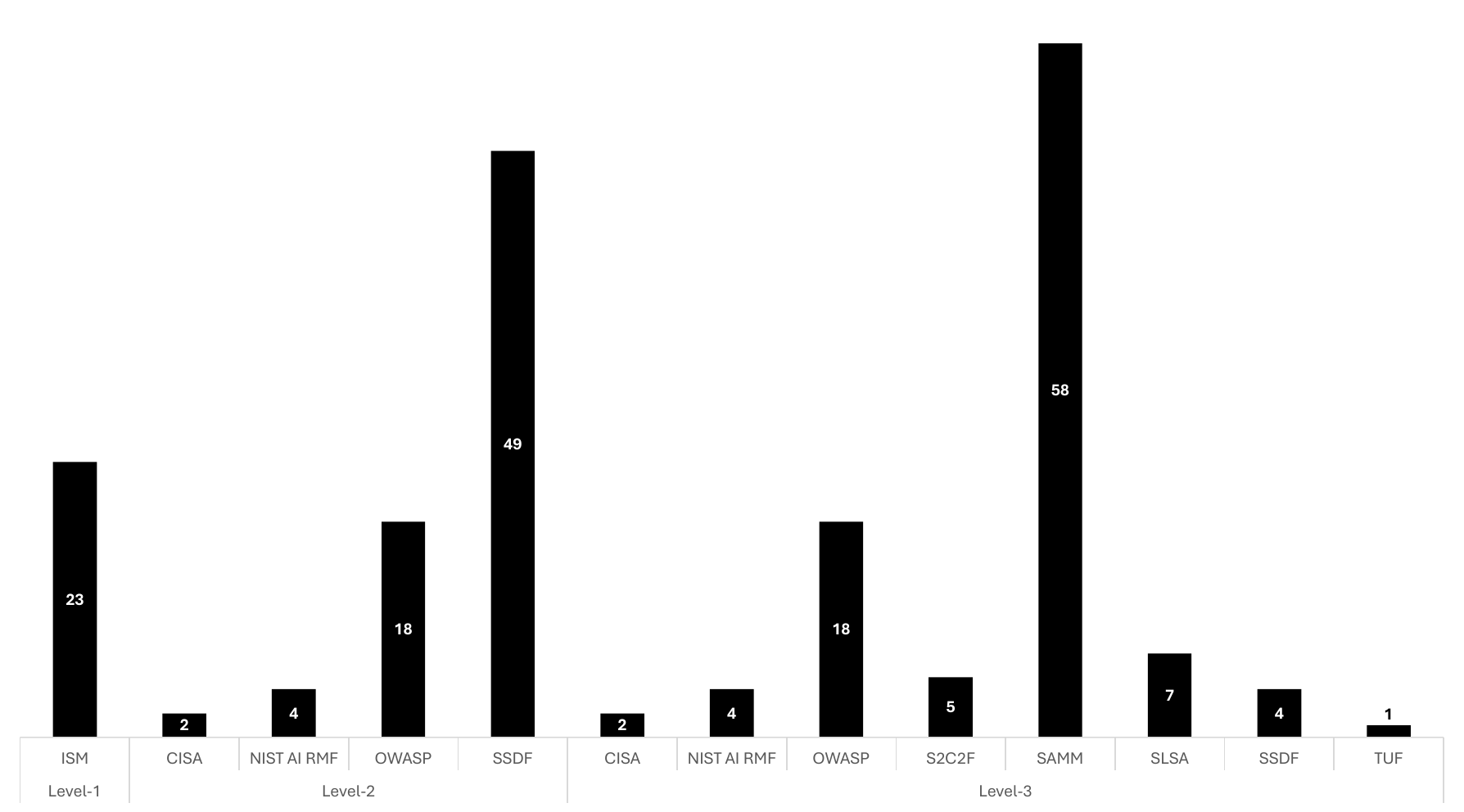}
    \caption{The distribution of requirements categorized by source frameworks, showcasing the contribution of each framework at different levels of granularity (Level-1, Level-2, Level-3).}
    \label{fig:statistic2}        
    \end{subfigure}
    \caption{Overview of the distribution of software security requirements and operations.\label{fig:statistic}}
\end{figure}

Figure \ref{fig:statistic} provides a detailed breakdown of the distribution of software security requirements across the defined goals and source frameworks used in this study. Figure \ref{fig:statistic1} shows how requirements and operations are allocated across the four software security goals. Each goal includes requirements at different levels (Level-1, Level-2, and Level-3), as well as associated operations, with the total counts highlighted. This figure highlights the prominence of the "Secure software development" goal, which contains the largest share of requirements and operations.

Figure \ref{fig:statistic2} presents the distribution of requirements based on their source frameworks, such as ISM, CISA, NIST AI RMF, SSDF, and SLSA. It shows how different frameworks contribute to requirements at varying levels of granularity (Level-1, Level-2, and Level-3). For example, ISM and NIST SSDF contribute significantly to higher-level requirements (Level-1 and Level-2, respectively), whereas SAMM and other frameworks provide more fine-grained details at Level-3. 

%The results were compared with industry benchmarks and related works in research such as SAMM, CISA and NIST to ensure the applicability and relevance of the mapping in real-world scenarios. The comparison result has been shown in Section \ref{sec:background}.

\textbf{Evaluation/Feedback:}
%\sunny{We may need feedback from others (internal/Google) for the next steps.}
The final phase involved reviewing the research outputs and collecting feedback from stakeholders, including external reviewers. 
We have shared the initial results with internal and external reviewers for feedback and improve the mapping.
Insights from the feedback were incorporated into the research, and a comprehensive final report was prepared, summarizing key findings and potential next steps.

The key feedback is on the usability and interoperability of the mapping. 

First, the mapping includes a large volume of requirements distributed across multiple layers, which can make it challenging for users to fully understand and navigate.
To address this limitation and improve accessibility and usability, we developed a web-based tool that allows users to explore the mapping more effectively. A detailed description and screenshots of the tool can be found in Section \ref{sec:webtool}.

Second, to support interoperability and machine-readability, we adopted the Open Security Controls Assessment Language (OSCAL), a standard format introduced by NIST. Our mapping format is aligned with OSCAL models, particularly the catalog and profile models used in the control layer. The data format and its application are explained in detail in Section \ref{sec:oscal}.

\textbf{Project Meetings and Timeline:}
Throughout the project, regular meetings were held to ensure consistent progress and alignment across all stakeholders. Weekly internal research team meetings focused on progress updates and issue resolution, while bi-weekly project team meetings involved all participants to discuss key updates, challenges and questions, and decision-making. The timeline for the research spanned from March 2024 to March 2025, reflecting a structured and iterative process to achieve the research objectives.

\section{Software Security Mapping Framework} \label{sec:Mapping}

\subsection{Framework Design}
We had seven steps to design and develop this framework. The following describes the detailed steps we conducted in this study.

\textbf{Define Top-most Strategic Goal:}
Defining a strategic goal is essential to ensure that this study addresses current problems and capitalizes on opportunities identified in collaboration with industry practitioners. 
This is a good starting point to provide a strong foundation for subsequent phases of the research.
In this study, we defined \textbf{\textit{"Secure Software"}} as the high-level strategic goal. This goal emphasizes the development of a comprehensive framework that mitigates existing security issues while promoting proactive practices to seize new opportunities in software security.

\textbf{Define Security Levels based on Focus and Depth:}
Under the high-level goal, we defined three requirement levels—Level-1, Level-2, and Level-3—based on their focus (e.g., strategic, operational, or technical) and depth of detail.

Level-1 focuses on strategic, high-level objectives without providing specific operational instructions. This level typically includes regulatory requirements, which are often described broadly without detailed technical specifications.

Level-2 provides a mid-level focus, offering general guidance with some details, but lacking the in-depth technical instructions needed for direct implementation.

Level-3 is the most detailed, focusing on technical-level requirements. It specifies how processes or requirements should be implemented to achieve the goals and meet higher-level requirements.

\textbf{Select and Review Existing Frameworks:}
This step involves conducting an in-depth literature review of both academic research and industry frameworks to survey and select appropriate frameworks for each requirement level. 
Based on this review, Australian Information Security Manual (ISM), 
NIST Secure Software Development Framework (SSDF), and the Supply-chain Levels for Software Artifacts (SLSA) and the Update Framework (TUF) were chosen as the primary frameworks for Level-1, Level-2, and Level-3, respectively.

\textbf{Identify Software Security Goals:}
In this step, we identified four software security goals that align with the strategic goal of "Secure Software". 
These goals were derived from the Level-1 requirements and include: "Secure Software Environment", "Secure Software Development", "Software Traceability", and "Vulnerability Management".

Section \ref{sec:goal} provides a detailed description of each goal.

\textbf{Elicit Requirements:}
We collected data (requirements) from the selected frameworks and categorized them into the four primary goals. These requirements were aligned with the high-level strategic goal of "Secure Software" and further refined using the KAOS approach to ensure comprehensive coverage of each goal.
Based on extensive data collection and analysis, we established a knowledge base that primarily aggregates requirements in their original form. We then added value by providing detailed explanations, structured interpretations, and practical insights. 

\textbf{Mapping: Group and Link Requirements:}
This step involved iteratively grouping and linking requirements across different levels to create a comprehensive mapping from Level-1 to Level-3.

First, requirements were mapped at a high level by aligning Level-1 requirements with the four defined goals. The primary objective was to establish broad categories for strategic alignment.

We then focused on linking Level-2 requirements to Level-1 by identifying mid-level guidance that supports strategic objectives. This round emphasized interpreting general guidance while maintaining consistency across levels.

Finally, Level-3 requirements were mapped to Level-2 by specifying detailed technical and operational requirements. These finer-grained requirements offered actionable details on how to implement the mid-level guidance.

We employed a "Traceability Matrix" as a tool for this mapping to ensure that high-level goals and requirements were systematically linked to lower-level implementation practices. The matrix traced relationships between goals, requirements, and final implementation steps, ensuring consistency, traceability, and completeness throughout the process. Thorough reviews were conducted during each round, and the resulting mapping was cross-checked by the research team. For exmaple, both top-down and bottom-up analyses were performed to identify potential gaps or missing requirements, ensuring no critical controls or practices were overlooked. This dual analysis approach enhanced the comprehensiveness and coverage of the framework.

\textbf{Identify Operations and Agents:}
Following the completion of the requirements mapping, we identified operations and agents for each requirement.
Operations can be identified using two different methods: one through stakeholder engagement (e.g., workshops), and the other by deriving them directly from requirements \cite{uszokkaos}. In this study, we first applied the latter approach and subsequently reviewed the results with relevant stakeholders.
Operations were derived from Level-3 requirements and represent specific and practical actions needed to achieve the goals and fulfill the requirements.
For each operation, agents, which include both stakeholders (e.g., software developers, security teams) and systems (e.g., automated security tools), were identified as responsible entities for implementing the operations and ensuring goal fulfillment.

\subsection{Overview of the Structure}

%The mapping framework was developed to align high-level security requirements with low-level requirements, ensuring that strategic objectives are effectively translated into actionable, detailed technical requirements. This approach provides a clear linkage between overarching goals and operational implementations, helping stakeholders navigate complex security requirements across different levels.

%To achieve coherence and consistency throughout the software supply chain, we created a traceability matrix that links high-level, mid-level, and low-level requirements. This traceability ensures that all phases, from strategic goal setting to technical implementation, are systematically connected. Both top-down and bottom-up analyses were performed to identify potential gaps or missing requirements, ensuring no critical controls or practices were overlooked. This dual analysis approach enhanced the comprehensiveness and coverage of the framework.

Our framework aims to facilitate stakeholder understanding by offering a structured, transparent process. It helps stakeholders clearly see how security requirements are implemented and understand their respective roles in fulfilling these requirements.

\begin{figure*} [htb]
    \centering
    \includegraphics[width=\textwidth]{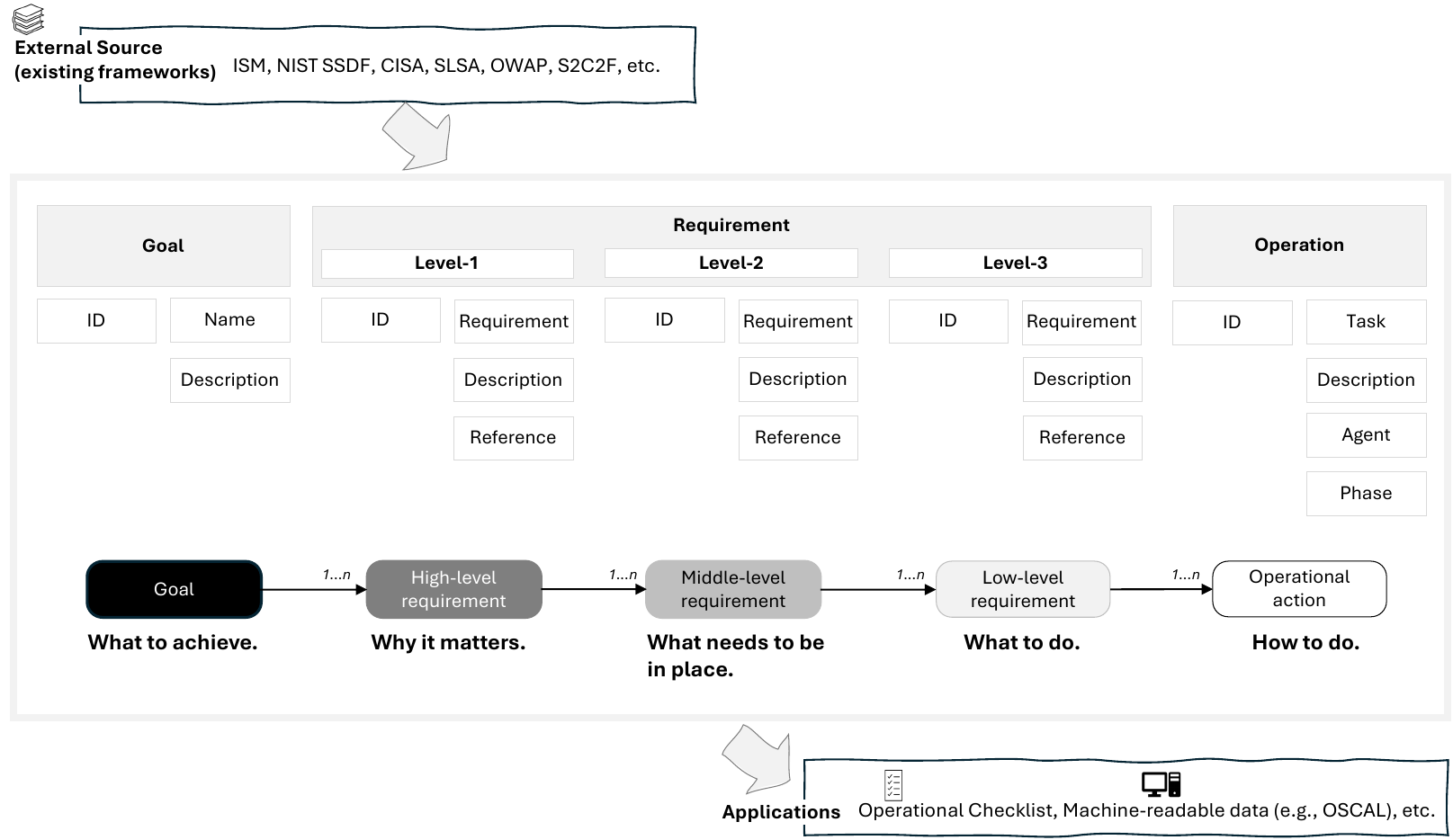}
    \caption{Overview of the software security mapping framework.}
    \label{fig:overview}
\end{figure*}

Figure \ref{fig:overview} illustrates an architecture of the mapping framework. The figure highlights the external sources used in developing the framework and potential practical applications, such as generating security checklists.

This improves stakeholder understanding by simplifying complex requirements and making them ready for day-to-day operational use. Organizations can easily generate operational checklists by extracting relevant requirements or directly selecting operations tailored to their specific security needs. This flexibility makes the framework a valuable tool for improving operational security practices in real-world scenarios.

\begin{figure*} [htb]
    \centering
    \includegraphics[width=0.9\textwidth]{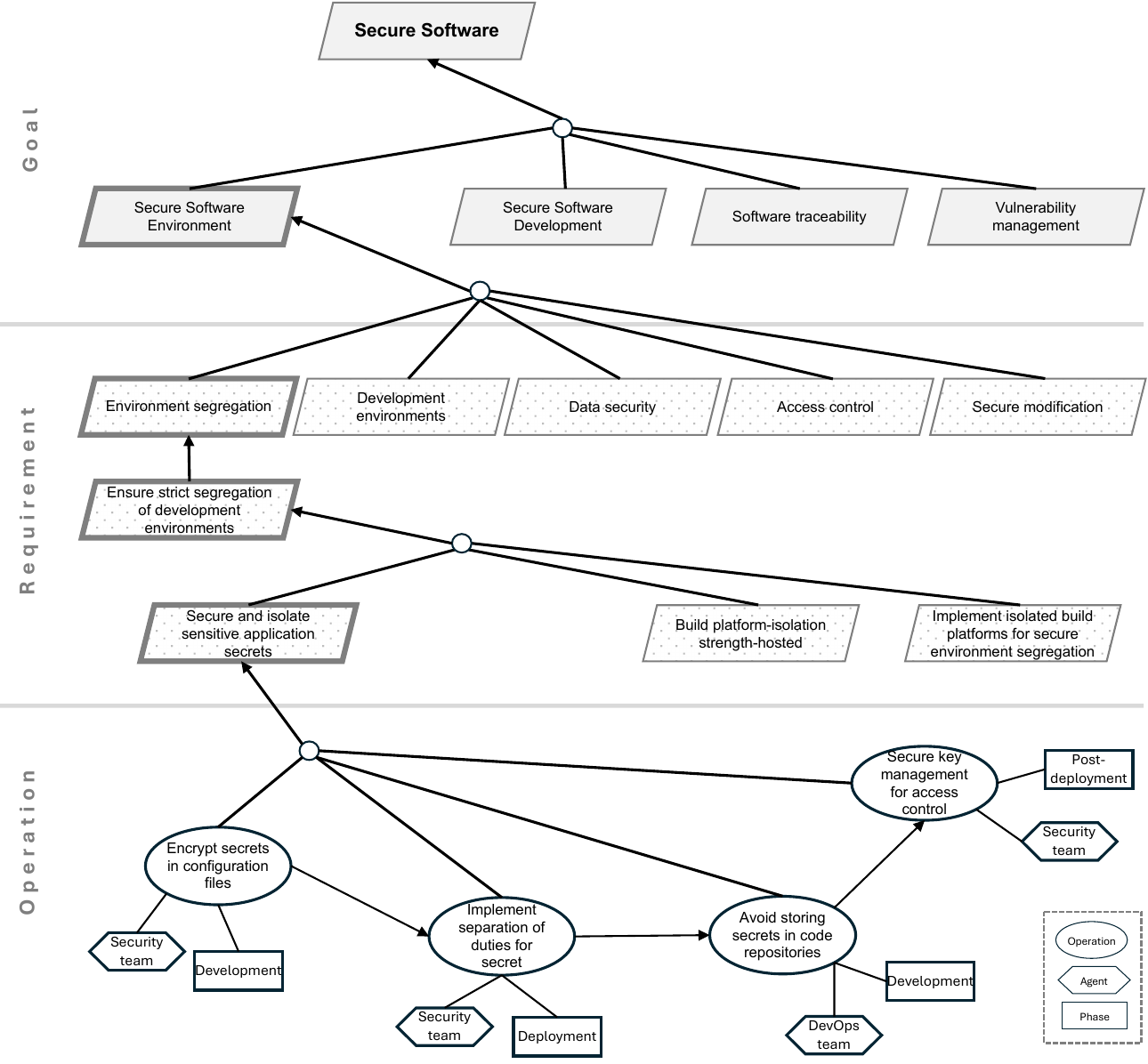}
    \caption{Example mapping between goal, requirement, and operation.}
    \label{fig:example}
\end{figure*}

Figure \ref{fig:example} presents an example mapping between goals, requirements, and operations.
The figure illustrates the top-most goal, its sub-goals, and five Level-1 requirements under the primary goal, "Secure Software Environment". 
The first Level-1 requirement, "Environment segregation", includes a Level-2 requirement that is further broken down into three Level-3 requirements: "Secure and isolate sensitive application secrets",  "Build platform-isolation strength-hosted", and "Implement isolated build platforms for secure environment segregation".
Additionally, the figure shows four operations associated with the first Level-3 requirement, which includes agents and software supply-chain phases.

In the following sections, we provide a detailed introduction to the mapping elements.

\subsection{Software Security Goals} \label{sec:goal}

As shown in Figure \ref{fig:example}, we identified "Secure Software" as the top-level strategic goal for this study.
This goal is supported by four sub-goals: "Secure Environment", "Secure Development", "Software Traceability", and "Vulnerability Management".

\begin{comment}
\begin{figure*} [htb]
    \centering
    \includegraphics[width=0.8\textwidth]{Fig/Goalmodel.pdf}
    \caption{The goal model of the framework; it includes the top-most strategic goal (Secure Software) and four goals to achieve the goal.}
    \label{fig:goalmodel}
\end{figure*}    
\end{comment}

\textit{Secure software environment} ensures that the infrastructure used for development, testing, and production is protected from unauthorized access, malicious code, and other security threats. This reduces the risk of vulnerabilities being introduced into the software and mitigates potential damage from security breaches \cite{goldberg1996secure}.

\textit{Secure software development} focuses on designing, developing, and testing software with security as a primary consideration. This proactive approach minimizes vulnerabilities and coding errors, reducing future security risks and enhancing the overall integrity of the software \cite{apvrille2005secure}.

\textit{Software Traceability} ensures that all components of the software, including their origins and any modifications, are documented and tracked \cite{cleland2014software}. This supports transparency, facilitates the identification of vulnerabilities, improves incident response, and ensures compliance with security and regulatory requirements \cite{bi2024way}.

\textit{Vulnerability Management} aims to identify, report, and resolve vulnerabilities promptly \cite{foreman2019vulnerability}. By implementing a vulnerability disclosure program and clear reporting mechanisms, organizations can address security risks quickly and reduce the likelihood of exploitation.

These four sub-goals collectively ensure that the overarching objective of "Secure Software" is achieved by addressing key aspects of security throughout the software life-cycle. By systematically implementing these goals, organizations can enhance software reliability, mitigate risks, and maintain compliance with security standards.

\subsection{Multi-layered Requirement}

\textit{Level-1 Requirement:}

This level comprises key components such as id, requirement, and reference including the source framework and relevant links. 
We identified 23 high-level requirements from the ISM framework, which primarily include regulatory requirements for various types of software development, such as traditional software, mobile software, and Large Language Models (LLMs). 
These high-level requirements serve as strategic objectives, ensuring compliance with industry regulations and standards.

Table \ref{tab:level1} presents the full list of the requirements we identified, categorized into the four goals. %The full list of Level-1 requirements can be found in Figure \ref{fig:level2}.

\begin{table}
    \centering
    \footnotesize
    \begin{tabular}{p{0.17\textwidth}p{0.25\textwidth}p{0.5\textwidth}}
    \hline
    Goal & Requirement & Description \\ 
    \hline
    Secure environment & Environment segregation & Development, testing and production environments are segregated. \\
    & Development environments	& Development and modification of software only takes place in development environments. \\
    & Data security & Data from production environments is not used in a development or testing environment unless the environment is secured to the same level as the production environment. \\
    & Access control & Unauthorised access to the authoritative source for software is prevented. \\
    & Secure modification & Unauthorised modification of the authoritative source for software is prevented. \\
    
    Secure development & Secure-by-design	& Secure-by-design and secure-by-default principles, use of memory-safe programming languages where possible, and secure programming practices are used as part of application development. \\
    & SecDevOps practices	& SecDevOps practices are used for application development. \\
    & Threat modelling & Threat modelling is used in support of application development. \\
    & Mobile security & The Open Worldwide Application Security Project (OWASP) mobile application security verification standard is used in the development of mobile applications.\\
    & LLM risk mitigation & The OWASP Top 10 for Large Language Model Applications are mitigated in the development of large language model applications. \\
    & Evaluation of LLM applications & Large language model applications evaluate the sentence perplexity of user prompts to detect and mitigate adversarial suffixes designed to assist in the generation of sensitive or harmful content. \\
    & Contents protection & Files containing executable content are digitally signed as part of application development. \\   
    & Secure configuration	& Secure configuration guidance is produced as part of application development. \\
    & Security testing & Applications are comprehensively tested for vulnerabilities, using static application security testing and dynamic application security testing, prior to their initial release and any subsequent releases. \\  
    
    Software traceability & Software bill of materials	& A software bill of materials is produced and made available to consumers of software. \\

    Vulnerability management & Vulnerability disclosure program	& A vulnerability disclosure program is implemented to assist with the secure development and maintenance of products and services. \\
    & Vulnerability disclosure policy & A vulnerability disclosure policy is developed, implemented and maintained. \\
    & Vulnerability disclosure process & Vulnerability disclosure processes, and supporting vulnerability disclosure procedures, are developed, implemented and maintained. \\
    & Security information	& A ‘security.txt’ file is hosted for all internet-facing organisational domains to assist in the responsible disclosure of vulnerabilities in an organisation’s products and services. \\
    & Vulnerability disclosure & Vulnerabilities identified in applications are publicly disclosed (where appropriate to do so) by software developers in a timely manner. \\
    & Vulnerability resolution & Vulnerabilities identified in applications are resolved by software developers in a timely manner. \\
    & Root cause analysis & In resolving vulnerabilities, software developers perform root cause analysis and, to the greatest extent possible, seek to remediate entire vulnerability classes. \\
    \hline
    \end{tabular}
    \caption{Level-1 requirements: 23 regulatory requirements which are derived from Australia Information Security Manual (ISM).}
    \label{tab:level1}
    \normalfont
\end{table}

\textit{Level-2 Requirement:}

At this level, each Level-1 requirement has been broken down into one or more mid-level requirements to provide more detailed guidance and operational context. 
These mid-level requirements offer clarity on how high-level objectives can be met in practice. 
A total of 73 requirements were identified, primarily based on the NIST SSDF framework. To ensure alignment with different goals and higher-level requirements, some of these mid-level requirements include intentional duplications, where the same requirement may apply across multiple categories or goals.

\begin{figure*}
    \centering
    \includegraphics[width=0.9\textwidth]{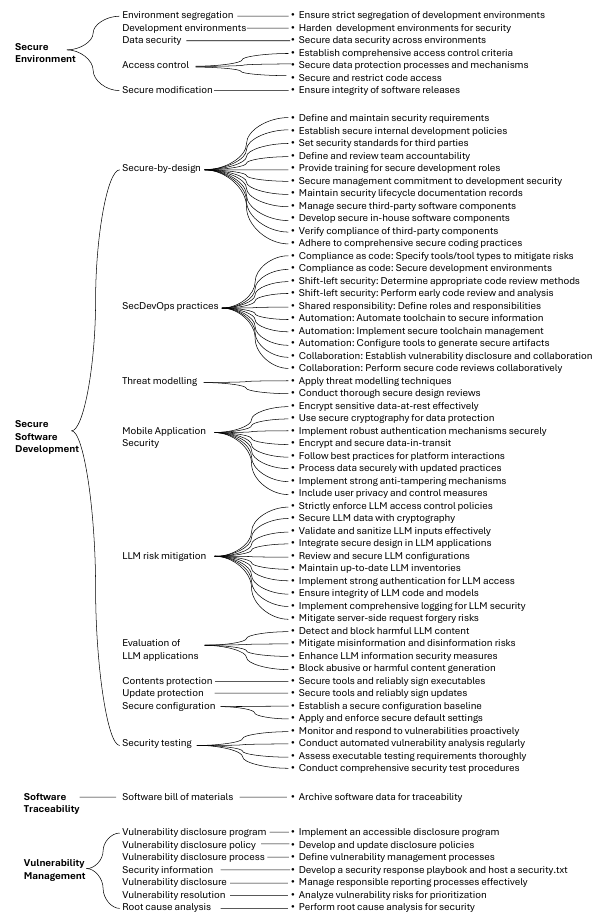}
    \caption{Alignment of Level-2 requirements with higher goals and requirements.}
    \label{fig:level2}
\end{figure*}

Figure \ref{fig:level2} presents all requirements at this level and shows how they align with the higher-level goals and requirements.

The goal of "Secure software environment" is to ensure that the infrastructure used for software development, testing, and production is protected against unauthorized access, data breaches, and other security threats. The primary focus is on maintaining a secure environment where software can be developed without introducing vulnerabilities.
At Level-2, this goal is operationalized by implementing specific requirements such as ensuring strict segregation between different environments (e.g., development, testing, production), securing data through encryption and robust access control mechanisms, hardening the development environment by applying secure configurations and regularly monitoring for threats, and establishing clear criteria for controlling and restricting access to critical assets and environments.

"Secure software development" is to ensure that software is developed with security as a primary consideration, minimizing vulnerabilities and maintaining the integrity of the software throughout its life-cycle.
This goal is achieved by defining and enforcing secure development policies and practices, establishing accountability by defining team roles and responsibilities, incorporating SecDevOps practices to integrate security into the development pipeline, applying threat modeling techniques and conducting secure design reviews to identify and mitigate risks early, ensuring compliance with secure coding standards and automating security checks (e.g., code review, testing), and providing training to development teams to enhance their understanding of secure development practices.

"Software traceability" is to ensure that all components of the software, including their origins and any changes, are documented and tracked. This enhances transparency and accountability, making it easier to identify vulnerabilities and ensure compliance with regulatory standards.
This goal can be achieved by maintaining a Software Bill of Materials (SBOM) that lists all software components, including third-party dependencies, and by archiving software data to ensure historical traceability and auditability. Additionally, ensuring that modifications to the software are traceable and documented enables quick identification of issues and facilitates effective incident response \cite{o2023impacts}.

The goal of "Vulnerability management" is to establish processes for identifying, reporting, and addressing vulnerabilities in a timely manner to reduce security risks.
This goal is implemented through a vulnerability disclosure program that enables responsible reporting of security issues and defines clear processes for disclosure and resolution. It involves developing a security response playbook, hosting a security.txt file to guide external reporters, and performing root cause analysis to address underlying issues and prevent similar vulnerabilities in the future. Additionally, regular automated vulnerability analysis and proactive monitoring help detect potential threats before exploitation \cite{ghaffarian2017software}.

\textit{Level-3 Requirement:}

Level-3 requirements were identified to provide finer-grained details, offering more precise technical and practical descriptions or examples. These include 99 requirements, primarily derived from SLSA, SAMM, and S2C2F, with additional consideration given to CISA, NIST AI RMF, and OWASP to complement the framework.

For example, the Level-2 requirement "Ensure strict segregation of development environments" is mapped to the Level-3 requirement "Build platform-isolation strength—hosted", which is based on the SLSA framework. This Level-3 requirement offers a detailed description along with specific examples as follows.

\textit{"All build processes must be executed on isolated, hosted build platforms operating on shared or dedicated infrastructure rather than individual workstations. This requirement enforces strict environmental segregation, aligning with the principles of zero-trust architecture and robust environmental protection. Examples of hosted build platforms include GitHub Actions, Google Cloud Build, and Travis CI."}

Each requirement at this level provides a similar level of detail, serving as a source for defining operations.
%The full list of Level-3 requirements can be found in a separated storage location \footnote{https://sunnylee.pythonanywhere.com/}.

\subsection{Actionable Operations}

Operationalization of security requirements supports clear responsibility and task allocation. It includes defining operations (tasks) and identifying agents responsible for executing these tasks to achieve the goals.
We derived operations from the Level-3 requirements, along with corresponding agents and relevant software supply-chain phases for each operation. Each Level-3 requirement generated 3 to 5 operations, resulting in a total of 424 operations in the framework.

Table \ref{tab:operation} shows example operations identified from "Implement isolated build platforms for secure environment segregation", Level-3 requirement. 
This requirement provides detailed description such as \textit{"Build platforms must enforce isolation to ensure that runs cannot influence each other, even within the same project. Isolation safeguards include preventing builds from accessing platform secrets, such as provenance signing keys, ensuring the integrity of the build provenance, ensuring builds that overlap in time cannot interact  preventing issues such as memory manipulation across builds, provisioning ephemeral environments for each build preventing persistent influences between consecutive builds, mitigating risks of "cache poisoning" by ensuring outputs remain consistent regardless of cache usage, and restricting remote influence or interactions unless they are explicitly captured and audited as external parameters."} in the framework.

We identified six operations from the requirement as shown in the table.

\begin{table} [htb]
    \centering
    \footnotesize
    \begin{tabular}{p{0.15\textwidth}p{0.2\textwidth}p{0.3\textwidth}p{0.1\textwidth}p{0.1\textwidth}}
    \hline
    Level-3 & \multicolumn{2}{l}{Operation} & Agent & Phase \\ 
    \hline
    Implement isolated build platforms for secure environment segregation & Ensure Secret Isolation & Configure the build platform to prevent builds from accessing platform secrets, such as provenance signing keys, to maintain the authenticity of the build provenance. This step ensures that secrets are fully isolated and inaccessible to build processes. & Development & Security Teams, Build Platform Engineers \\
    
    & Enforce ephemeral build environments & Provision an ephemeral, isolated environment for each build to prevent one build from persisting data or influencing subsequent builds. This guarantees that each build runs independently and does not retain any state after completion. &	Deployment	& DevOps Teams, IT Operations \\

    & Prevent concurrent build interference	& Implement controls to ensure that builds running concurrently cannot affect one another. For example, prevent any shared memory access or other forms of cross-build influence to ensure each build is fully isolated, even if they overlap in time.	& Development	& Infrastructure Teams, Security Engineers \\

    & Safeguard against cache poisoning	& Configure the build platform to prevent one build from injecting false entries into a shared build cache used by another build, ensuring consistent build outputs regardless of cache usage. This step protects against "cache poisoning" attacks.	& Post-deployment	& DevOps Teams, Security Analysts \\

    & Restrict remote influence	& Ensure the build platform does not open services that allow remote influence on the build environment unless explicitly defined as external parameters in the provenance. This step prevents unauthorized remote access or control over the build process.	& Development	& Security Teams, Build Platform Engineers \\
    
    \hline
    \end{tabular}
    \caption{Example operations identified from Level-3 requirement.}
    \label{tab:operation}
    \normalfont
\end{table}

The operations are detailed alongside the responsible parties ("Agent") and the timeline ("Phase") in which they are conducted. In the table, "Agent" identifies the team or individual accountable for specific tasks, while "Phase" categorizes the stage of the software supply chain where the operations occur.

As the framework was designed for general-purpose use and to support a wide range of applications, the software supply chain is divided into four key phases: preparation, development, deployment, and post-deployment. General roles, such as the security team and development team, are also defined to ensure clarity and accountability.

This structure provides organizations with the flexibility to customize and adapt the framework to meet their unique operational needs and contextual requirements, enabling efficient and effective implementation of security practices across their software supply chain.

\subsection{Validation of the Framework} 

Our work builds on previous efforts to map software supply chain security frameworks (e.g., SSDF<->SAMM). 
While these mappings are valuable, they are primarily crosswalks between frameworks at similar abstraction levels and stop short of offering actionable and lack operational detail. 
They also do not provide a unified structure that links strategic intent to practical implementation, nor address traceability across the full software life-cycle.
In contrast, our framework introduces several novel contributions as follows.

\textbf{Multi-layered and goal-driven structure.}
We move beyond flat framework-to-framework mappings by applying goal-oriented requirements engineering to create a hierarchical, three-level mapping, from strategic (Level-1) to operational (Level-3) requirements. This structure provides semantic clarity and traceable linkages between goals, requirements, and implementation actions, addressing a key limitation of prior mappings that often lacked internal consistency or top-down rationale.

\textbf{Operationalization of requirements.}
A core contribution of our framework is the derivation of over 400 operations grounded in Level-3 requirements. These operations are detailed and context-aware tasks that specify who (agent) performs what (operation), when (phase), and why (linked requirement and goal). This operational focus addresses a well-recognized gap in existing frameworks, which tend to remain at the level of high-level policies or control objectives without specifying actionable steps.

\textbf{Enhanced interoperability.}
To support integration with existing tools and practices, we adopted and extended the NIST Open Security Controls Assessment Language (OSCAL). Our adaptation includes a new <operation> component within the OSCAL catalog model, allowing organizations to represent and share fine-grained operational practices in a machine-readable format (see Section \ref{sec:oscal} for details). This supports automation, toolchain integration, and streamlined compliance reporting, which is not addressed by prior mapping frameworks.

\textbf{Comprehensive framework and supply chain coverage.}
While previous mappings focus on aligning two frameworks (e.g., SSDF and SAMM), our approach unifies inputs from seven major frameworks (ISM, SSDF, SLSA, SAMM, TUF, CISA, S2C2F), enabling broader coverage of diverse software security concerns. Importantly, this mapping also comprehensively spans the entire software supply chain, from the early preparation phase through development, deployment, and post-deployment activities. Each operation is contextualized within its relevant phase and assigned to appropriate agents, ensuring that responsibilities and actions are clearly traceable across the full life-cycle. Through iterative review and refinement, we achieved 100\% mapping completion across all identified goals and requirements, demonstrating the robustness, extensibility, and practical applicability of our approach to real-world secure software development and supply chain security management.

\textbf{Real-world validation through incident-based checklist.}
We validated the practical utility of our framework by generating a scenario-based checklist in response to the \textit{Log4j vulnerability (CVE-2021-44228)} (see Section \ref{sec:webtool} for details). By extracting relevant operations tied to risk mitigation recommendations, we demonstrated how our framework can be used to rapidly assess, plan, and communicate security actions in real-world scenarios which existing mappings do not directly support.

In summary, our framework advances the field by offering operationally detailed and practically validated model that is tailored for end-to-end software supply chain security. It addresses not only what needs to be done (requirements), but also how it should be done (operations), by whom (agents), and when (phases), enabling a more complete and actionable understanding of secure software development practices.
We compared with key existing mappings as shown in Table \ref{tab:comparison}.

\begin{table*} [htb]
\footnotesize
\centering
\caption{Comparison with the existing mapping frameworks.}
\label{tab:comparison}
\rowcolors{2}{white}{gray!5}
\begin{tabular}{p{0.15\textwidth}p{0.2\textwidth}p{0.15\textwidth}p{0.17\textwidth}p{0.2\textwidth}}
\hline
Framework & Mapping structure & Operationalization & Machine-readability & Coverage \\
\hline
SAMM-SSDF & Flat: control-to-control & Not addressed & Not supported & \shortstack[l]{2 frameworks, \\ entire supply-chain} \\
SLSA-SSDF & Flat: control-to-control & Not addressed & Not supported & \shortstack[l]{2 frameworks, \\ deployment phase} \\
SDL-SAMM & Flat: control-to-control & Limited (examples) & Not supported & \shortstack[l]{2 frameworks, \\ entire supply-chain} \\
S2C2F-others & Flat: control-to-control & Not addressed & Not supported & \shortstack[l]{Multiple frameworks, \\ development phase} \\
\textbf{Our mapping} & \shortstack[l]{\textbf{Multi-layered:} \\ \textbf{goal-requirement-operation}}  & \textbf{Addressed} & \textbf{Supported (OSCAL)} & \shortstack[l]{\textbf{Multiple frameworks,} \\ \textbf{entire supply-chain}} \\

\hline

\end{tabular}
\end{table*}

\section{Discussion and Implications} \label{sec:Discussion}

\begin{comment}

\sunny{Thinking two options for this section:}

\subsection{Ecosystem???}
\sunny{We can discuss later if we include this section.}

\begin{itemize}
    \item Suggest a potential ecosystem (collaborations, participating groups, and impacts (?)
    \item Show a web application (screenshots) to support this ecosystem.
    \item Explain how this can be used: for an overview of software security requirements/operations in a user-friendly way.
    \item Explain how users can contribute to this (e.g., providing references) and what the impacts on the ecosystem are.
\end{itemize}

\sunny{We can mention challenges, gaps/limitations of this study and mention ecosystem as future directions at the end (or in the conclusion?).}

While this study provides comprehensive software security requirements and operational guidelines for practical application, it is acknowledged that these may not be exhaustive or flawless. 
Therefore, the framework (?) envisions a dynamic ecosystem that fosters self-growth, where multiple participating groups can interact, contribute, consume, and collaborate. This collaborative environment enables continuous refinement and expansion of security practices to address evolving challenges and emerging needs.

\end{comment}

\subsection{Real-world applications of the mapping} \label{sec:webtool}
This section discusses the potential applications of the mapping framework, including a user navigation tool and example checklists designed to mitigate potential software security risks for specific concerns and scenarios.

\begin{figure} [!]
    \centering
    \begin{subfigure}{\textwidth}
    \includegraphics[width=0.95\textwidth]{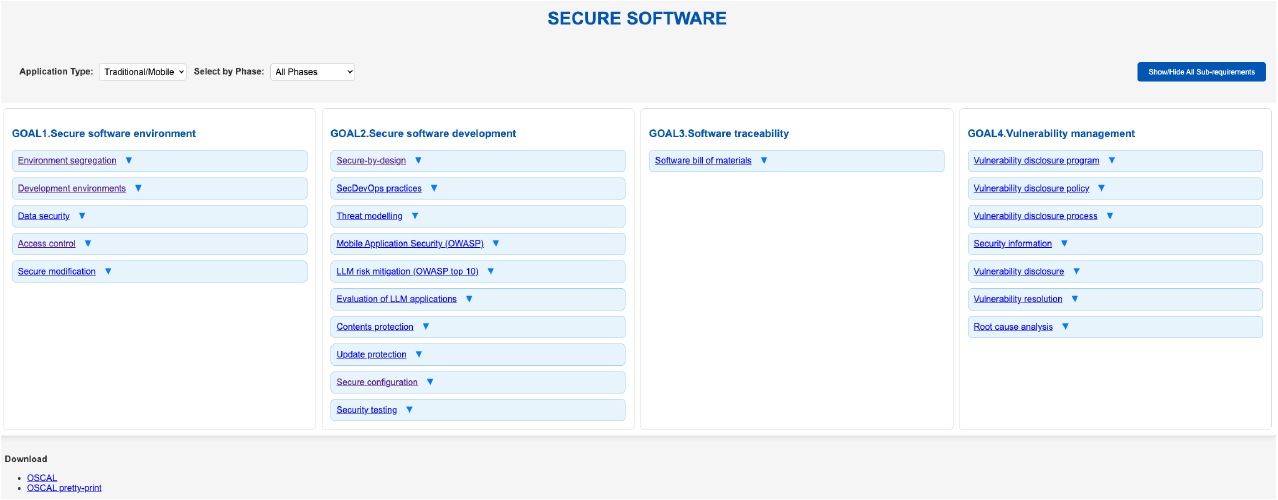}
    \caption{Navigator for the mapping; It shows four software security goals and level-1 requirements for each goal.}
    \label{fig:tool1}        
    \end{subfigure}
    \begin{subfigure}{\textwidth}
    \includegraphics[width=0.95\textwidth]{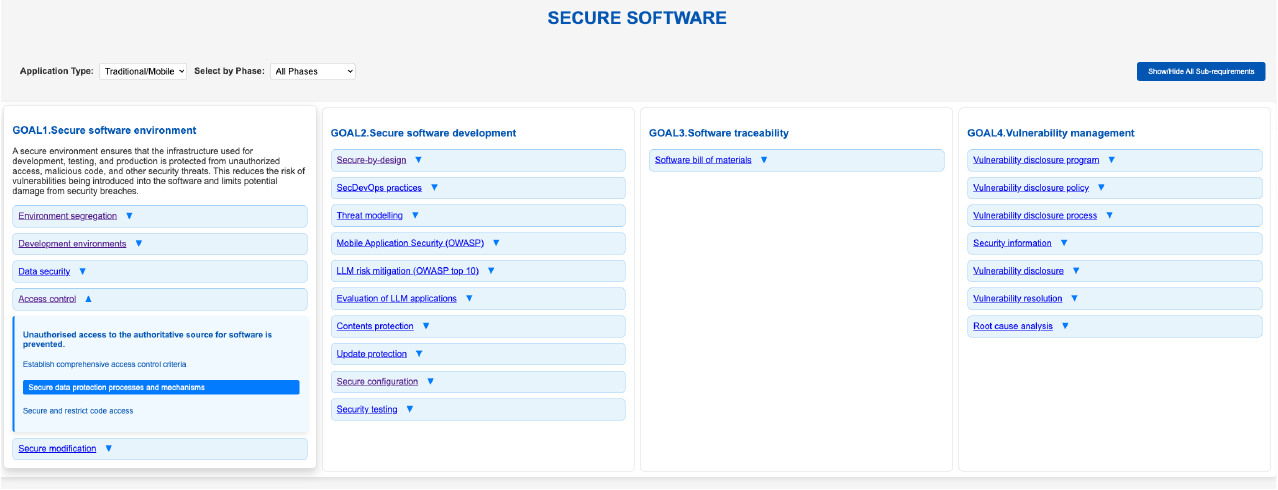}
    \caption{Selection of level-2 requirement; "Access control" has three level-2 requirements.}
    \label{fig:tool2}        
    \end{subfigure}
    \begin{subfigure}{\textwidth}
    \includegraphics[width=0.95\textwidth]{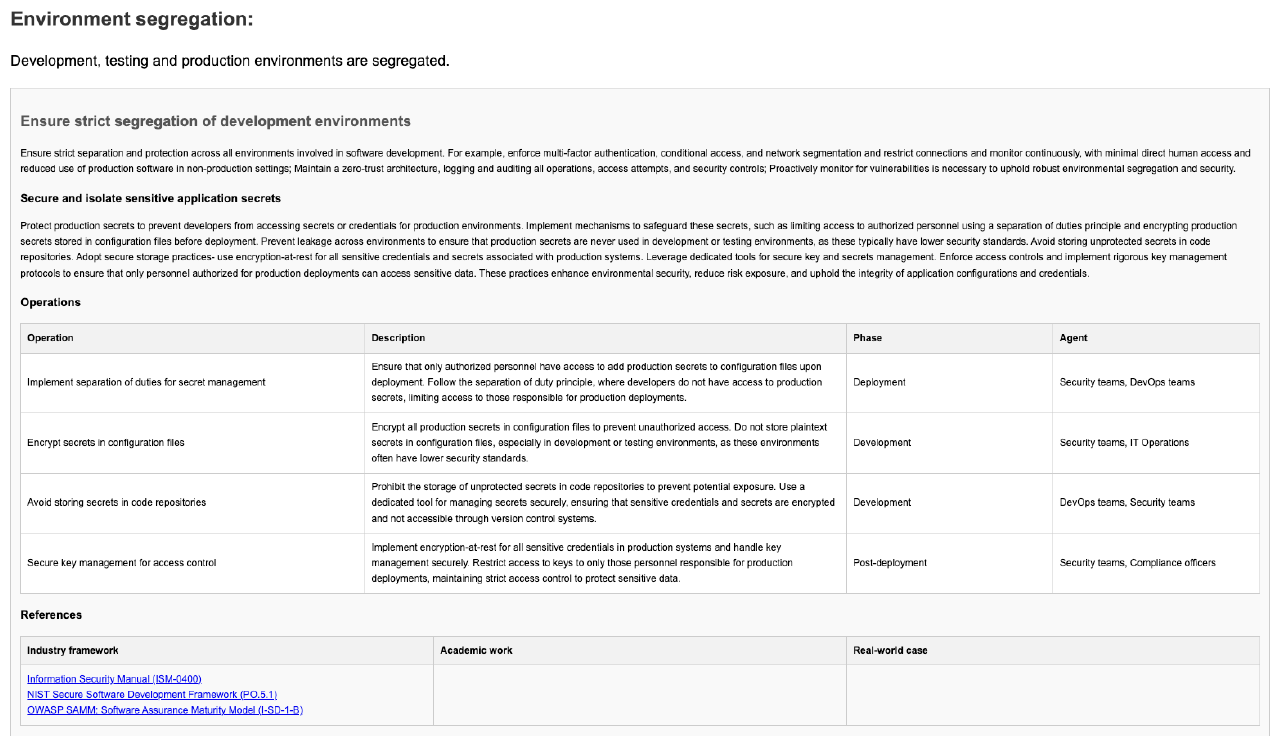}
    \caption{Level-3 and operations; A user can select Level-1 or Level-2 requirement and see the detailed requirements and operations.}
    \label{fig:tool3}        
    \end{subfigure}
    \caption{Web-based tool for exploring the mapping framework.} \label{fig:tool}
\end{figure}

\textbf{Interactive tool for mapping exploration:}
The mapping developed in this study encompasses a large number of requirements and operations under four overarching goals. Potential users, such as software developers and security experts within organizations, may face challenges in understanding the mapping due to its high volume and complex structure. 
To address this issue and enhance accessibility, we have developed a web-based navigation tool \footnote{https://sunnylee.pythonanywhere.com/} that provides a simple yet efficient way to explore the mapping (Figure \ref{fig:tool}).

Figure \ref{fig:tool1} shows the top-level strategic goal, "Secure Software," along with its four sub-goals and corresponding Level-1 requirements. 
Users can expand each goal to view its description before exploring the associated requirements in greater detail (Figure \ref{fig:tool2}).

For each requirement, the tool provides a brief description and a list of the next-level requirements (Level-2). 
Users can click on any requirement to navigate to the next level and see more detailed requirements and associated operations.

As shown in Figure \ref{fig:tool3}, selecting a Level-2 requirement leads to a page that presents its detailed requirements and operations. 
This page provides comprehensive information about the selected requirement, including the Level-3 requirements and operations linked to each of them.

While this study provides a comprehensive set of software security requirements and operational guidelines for practical application, it is recognized that these may not be exhaustive. This tool can serve as a platform to engage users and foster an ecosystem where multiple stakeholders can contribute to the mapping and its ongoing improvement. 
A dynamic, collaborative ecosystem can enable self-sustaining growth, allowing diverse groups to interact, contribute, utilize, and collaborate. 
Such an environment facilitates the continuous refinement and expansion of security practices to address emerging challenges and evolving needs.

\textbf{Scenario-based security checklists:}
To demonstrate the practical application of our mapping framework, we selected a real-world incident, \textit{"Log4j Vulnerabilities Create Unprecedented Impacts Worldwide"}. 
This case study is discussed in \textit{"CISA Tabletop Exercise Package Open Source Software"} and has been registered in the Common Vulnerabilities and Exposures (CVE) database and publicly visible with the assigned identifier, CVE-2021-44228 \cite{MitreCVE}. The following is the case description from the CISA report.

\begin{quotation}
In November 2021, critical vulnerabilities were discovered in a widely used, open source Java-based logging framework. The vulnerability set allowed for remote code execution that could be exploited in Java installations worldwide. The vulnerabilities became a zero-day exploit, with an upgraded version made publicly available the day after the first exploit was observed. Mitigation and remediation required unprecedented levels of effort from individual organizations and the broader cybersecurity community. The Java Naming and Directory Interface lookup feature, incorporated into Log4j in 2014, introduced the vulnerable attack surface. 
Log4j is considered an “endemic vulnerability” because vulnerable versions of Log4j will remain in systems for years to come. Organizations should have long-term capabilities to discover and upgrade vulnerable software to reduce the risks created by this endemic vulnerability, to include proactively monitoring for and upgrading vulnerable versions of Log4j, preventing the reintroduction of vulnerable versions of Log4j, and prioritizing applying software upgrades to avoid long-term exposure of vulnerable attack surfaces \cite{CISAOpen}.
\end{quotation}

To address the concerns raised by the Log4j incident, we identified 24 recommendations and practices for organizations from the review reports of this incident \cite{CISAOpen, CSRBLog4j} and mapped to the requirements and operational actions from our framework. 
The recommendations encompass actual organizational responses, lessons learned, and expert guidance discussed in the reports.

Table \ref{tab:checklist} presents the list of recommendations along with the corresponding security checklist (a total of 21) generated using our mapping framework.
The table also provides brief instructions for each checklist item. For detailed information and operational tasks, our mapping framework provides a structured approach that connects specific security requirements to actionable steps.

\begin{longtable}                   
    {p{0.25\textwidth}p{0.4\textwidth}p{0.27\textwidth}} 
    \hline
    Recommendation & Checklist & Instruction \\ 
    \hline
    \endfirsthead
    \hline
    Recommendation (continued) & Checklist & Instruction \\ 
    \hline
    \endhead
    \hline
    \multicolumn{3}{r}{\textit{Continued on next page}} \\ \hline
    \endfoot
    \endlastfoot
    \rowcolor{gray!5}1. Promote strong security hygiene practices. & 
    $\square$ Adhere to comprehensive secure coding practices. 
    & Promote adherence to coding standards to enhance overall software security. \\
    
    2. Enhance knowledge and awareness of the vulnerability. & $\square$ Provide training for secure development roles. & Enhance awareness through training and information sharing. \\

    3. Train stakeholders in secure software development. & & Equip developers with the skills needed for secure coding practices. \\

    \rowcolor{gray!5}4. Build a secure software ecosystem. & $\square$ Manage secure third-party software components. & Focus on the secure integration and management of third-party software. \\

    5. Implement egress filtering to manage network traffic from systems affected by the Log4j vulnerability. & Ensure strict segregation of development environments. & Ensure strict separation and monitor environments to prevent unauthorized data flow. \\
    & $\square$ Build platform-isolation strength-hosted. & Use isolated and secure platforms and ensure that unauthorized/potentially malicious traffic does not affect the system. \\
    & $\square$ Implement isolated build platforms for secure environment segregation. & Restrict unwanted outbound communication from potentially vulnerable systems. \\
    & $\square$ Secure and isolate sensitive application secrets. & Prevent unauthorized access to sensitive application secrets by blocking malicious traffic attempting to exploit known vulnerabilities like Log4j. \\

    \rowcolor{gray!5}6. Prevent potential attack by configuring Web Application Firewalls (WAF). & $\square$ Establish a secure configuration baseline. & Establish secure defaults and ensure that configurations support security functions without weakening protections. \\

    7. Develop and utilize internal and external communication channels effectively. & $\square$ Establish vulnerability disclosure and collaboration. & Facilitate effective communication and sharing of mitigation strategies among stakeholders. \\

    8. Recognize the security limitations of open source projects, which often lack dedicated security resources. & & Encourages collaboration to enhance security in open source projects. \\

    \rowcolor{gray!5}9. Disconnect affected assets from the network until they are upgraded. & $\square$ Conduct comprehensive security test procedures. & Isolate affected systems to prevent further compromise. \\

    10. Assess business risks, update vulnerable software, and monitor for malicious activity. & Monitor and respond to vulnerabilities proactively. & Enhance response times by actively monitoring for threats. \\
    11. Proactively monitor and upgrade vulnerable versions to prevent their reintroduction. & $\square$ Continuously monitor and remediate vulnerabilities in open-source software dependencies. & Ensure vulnerabilities are addressed quickly to minimize risk. \\
    12. Commit to long-term strategies for identifying and upgrading vulnerable software. & $\square$ Enforce timely patch management across the portfolio. & Ensure vulnerabilities are discovered and addressed in real-time to minimize risks. \\

    \rowcolor{gray!5}13. Improve SBOM tooling and adoption while promoting software transparency. & $\square$ Archive software data for traceability. & Enhance visibility into software supply chains. \\

    14. Track and manage the users of produced software and understand its eventual usage. & $\square$ Safeguard provenance data for transparency. & Improve traceability and understanding of software usage. \\

    \rowcolor{gray!5}15. Ensure access to technical resources and establish mature processes. & $\square$ Implement an accessible disclosure program. & Support transparent and responsible vulnerability disclosures. \\
    \rowcolor{gray!5}& $\square$ Develop and update disclosure policies. & Outline the procedures for identifying, reporting, and remediating vulnerabilities. \\
    \rowcolor{gray!5}& $\square$ Define vulnerability management processes. & Ensure organizations have defined processes to handle and remediate vulnerabilities efficiently. \\

    16. Implement existing incident response, crisis, or vulnerability management plans. & $\square$ Develop a security response playbook and host a security.txt. & Ensure organizations are prepared to handle and mitigate vulnerabilities efficiently. \\
    17. Maintain a documented vulnerability response program, including disclosure and handling procedures. & & Ensure a structured approach for vulnerability reporting, handling, and mitigation. \\

    \rowcolor{gray!5}18. Immediately disclose the vulnerability. & $\square$ Manage responsible reporting processes effectively. & Establish a central, organization-wide Product Security Incident Response Team (PSIRT) to manage vulnerability disclosures and remediation efforts. \\

    19. Ensure risk trade-offs in software usage and integration. & $\square$ Analyze vulnerability risks for prioritization. & Evaluate trade-offs to minimize risks from vulnerable software usage. \\
    20. Assess and manage the potential risk, at both a technical and business level (including ongoing risk management processes). & & Evaluate risks comprehensively to align technical and business considerations for mitigation. \\

    \rowcolor{gray!5}21. Apply patches or disable vulnerable code sections to mitigate risks. &  $\square$ Implement risk-based remediation plans. & Apply patches or mitigations to reduce the risk of exploitation. \\
    \rowcolor{gray!5}22. Prioritize software upgrades to minimize long-term exposure to vulnerable attack surfaces. & & Reduce long-term risks by prioritizing critical upgrades. \\
    \rowcolor{gray!5}23. Address continued risks of the vulnerability. & & Ensure ongoing monitoring and management of vulnerabilities to prevent future risks. \\

    24. Delete the JndiLookup.class file containing the vulnerable code. & $\square$ Eliminate vulnerability classes proactively. & Addresses vulnerabilities by removing affected components entirely. \\
    
    \hline
    \caption{Checklist generated using our mapping framework, incorporating recommendations from Log4j incident reports. \label{tab:checklist}}
\end{longtable}

\begin{figure} [htb]
    \centering
    \begin{subfigure}{0.6\textwidth}
    \includegraphics[width=\textwidth]{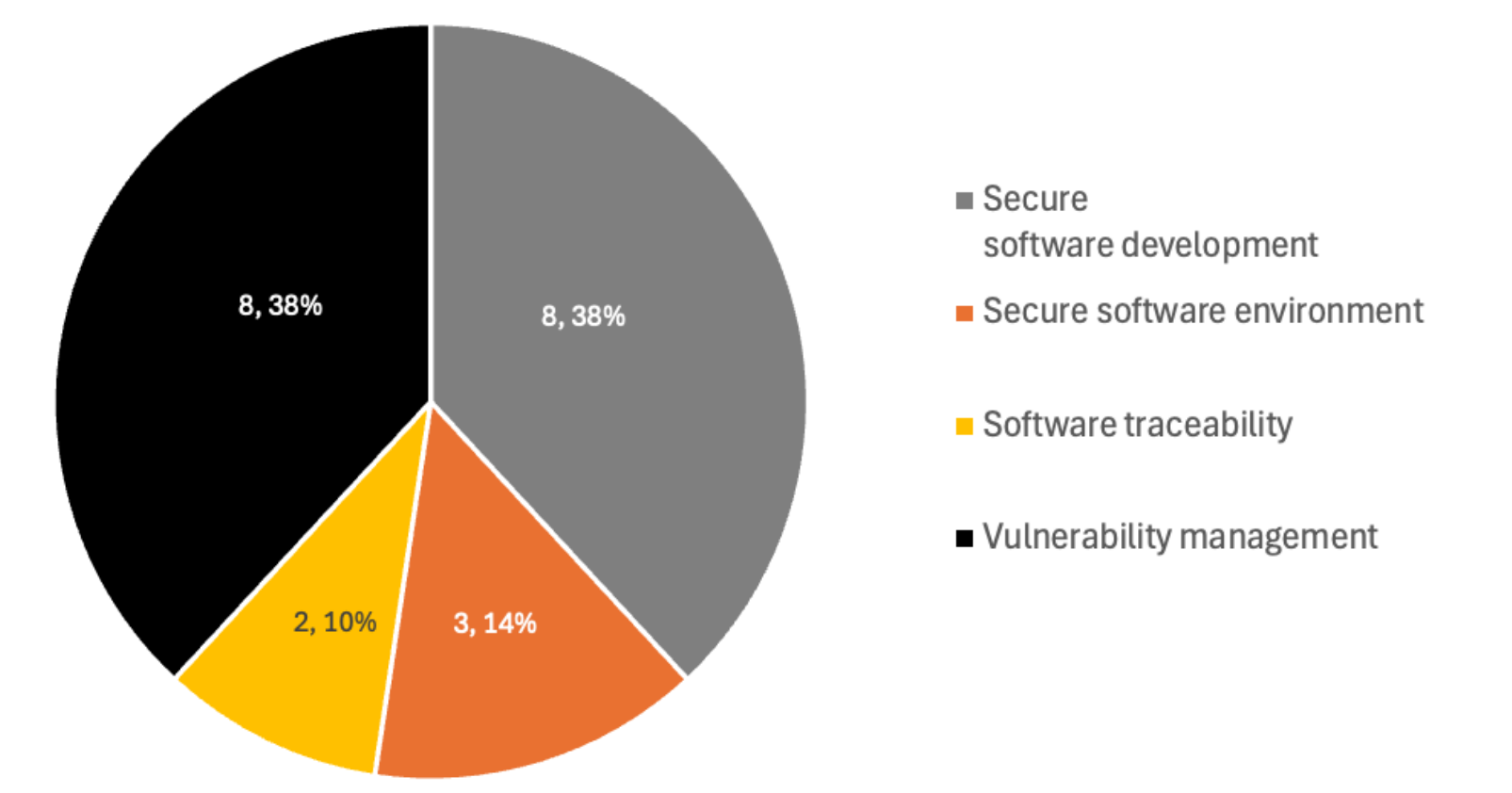}
    \caption{Checklist distribution by software security goals.}
    \label{fig:checklist1}        
    \end{subfigure}
    \begin{subfigure}{\textwidth}
    \includegraphics[width=\textwidth]{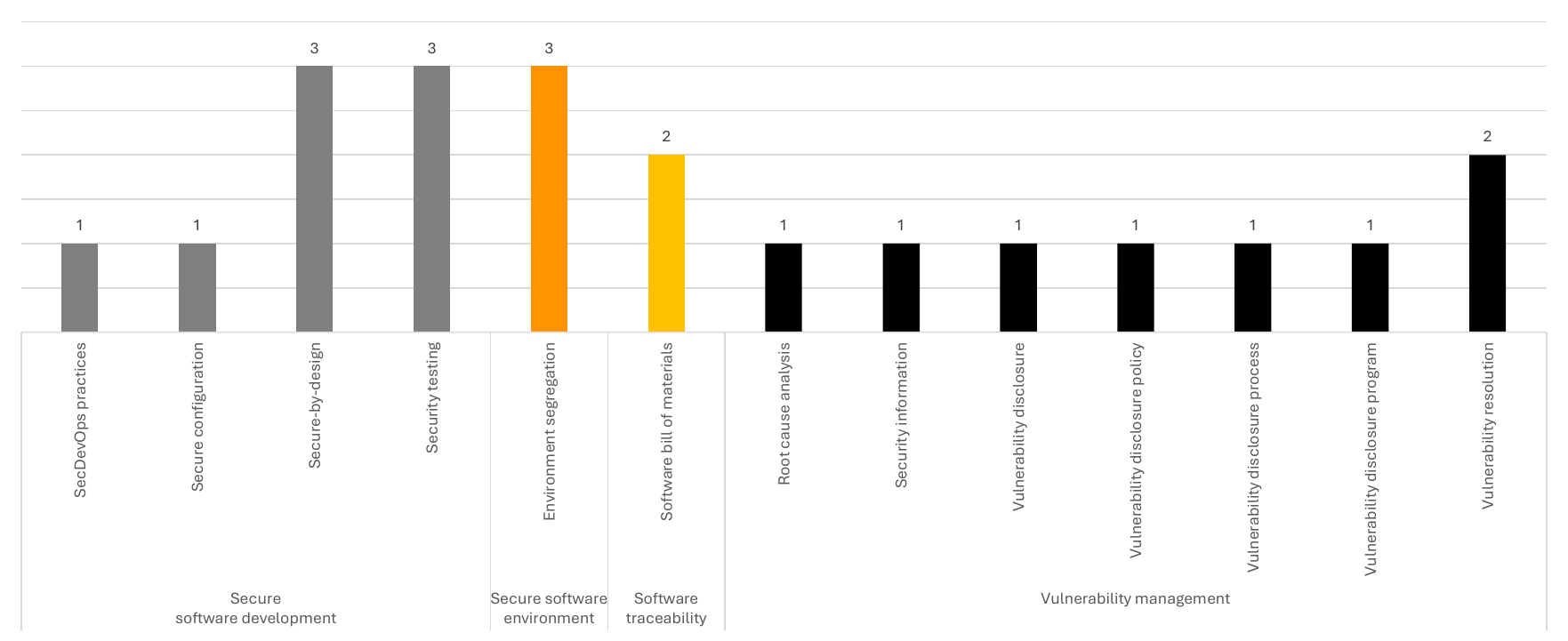}
    \caption{Breakdown of the checklist by software security goals and high-level categories.}
    \label{fig:checklist2}        
    \end{subfigure}
    \caption{Distribution and categorization of the checklist derived from the mapping framework. \label{fig:checklist}}
\end{figure}

Although the checklist was developed to address a specific security incident (Log4j), it comprehensively covers various aspects of software security. In addition to vulnerability management, it also includes areas such as secure software environments, secure development, and software traceability (Figure \ref{fig:checklist}).
Furthermore, the checklist can serve as a proactive security framework for organizations to strengthen their overall security posture and mitigate risks beyond Log4j-related threats.

Figure \ref{fig:checklist1} shows identified checklist distribution by software security goals. The figure illustrates the proportion of checklists categorized under the four software security goals. 
As shown, the majority of the checklist fall under "Secure Software Development (38\%)" and "Vulnerability Management (38\%)", reflecting their critical importance in addressing security concerns such as coding standards, vulnerability identification, and mitigation. 
"Secure Software Environment" and "Software Traceability" are comparatively less represented, with contributions of 14\% and 10\%, respectively, emphasizing their specialized but essential roles.

Figure \ref{fig:checklist2} is a breakdown of the checklist by software security requirements. It provides a detailed count of the checklist mapped to specific requirements (Level-1) within each software security goal.
"Secure Software Development" includes key checklists for SecDevOps practices, secure-by-design principles, and secure testing, with each category having significant contributions.
"Secure Software Environment" emphasizes environment segregation, which accounts for the highest number of checklists in this category.
"Software Traceability" primarily focuses on maintaining and safeguarding the software bill of materials (SBOM) to ensure transparency and traceability.
"Vulnerability Management" spans multiple high-level categories such as vulnerability disclosure programs, vulnerability resolution, and root cause analysis, highlighting its comprehensive approach to addressing and mitigating vulnerabilities.

\subsection{Machine-readable Format for Interoperability} \label{sec:oscal}

%\sunny{we need to think about which OSCAL models should be included in this section; Catalog model only, Catalog + Profile (from the previous section: checklist), or more models?}

Machine-readable formats such as Open Security Controls Assessment Language (OSCAL)\cite{OSCAL}, Structured Threat Information Expression (STIX)\cite{STIX}, Common Security Advisory Framework (CSAF)\cite{CSAF} are essential for modern cybersecurity, compliance, and risk management. They enable automation, accuracy, portability and interoperability, reducing manual effort, minimizing errors, and streamlining security assessments. 
In the context of software supply chain security, OSCAL plays a critical role in facilitating the structured representation, validation, and exchange of security requirements/controls across different organizations and tools.

To support a comprehensive and practical implementation of software supply chain security frameworks, we provide both an Excel sheet and a web tool for holistic mapping. This mapping spans three levels of frameworks down to detailed operational steps, offering a structured reference architecture model that is both practical and adaptable across diverse real-world scenarios.

While the human-readable representation of the Software Security Framework Mapping ensures stakeholder transparency and accessibility, its machine-readable specification, based on the OSCAL model, enhances portability and interoperability for documentation, reporting, and information sharing across the software supply chain community. OSCAL standardizes the representation of security controls, enabling security frameworks to define, search, import, and export control information in a unified format. This common structure ensures seamless integration with automated compliance tools, risk management systems, and DevSecOps pipelines, thereby strengthening the security posture of the software supply chain.
     
\subsubsection{Introduction to OSCAL}

OSCAL is a standardized framework developed by NIST to facilitate security assessment, authorization, and continuous monitoring of information systems. OSCAL provides a machine-readable format for security control-related data, supporting multiple formats such as JSON, YAML, and XML\cite{piez2019open}. By leveraging OSCAL, organizations can automate compliance processes, improve security documentation consistency, and enhance interoperability across security tools and frameworks.

OSCAL is structured into three layers including \textit{Control layer}, \textit{Implementation layer}, and \textit{Assessment layer} (Figure \ref{fig:catalog}). Each layer contains a set of models that support different aspects of security control implementation. 

    % \begin{figure*} [htb]
    % \centering
    % \includegraphics[width=0.6\textwidth]{Fig/OSCAL.drawio.pdf}
    % \caption{OCSAL Model Overview.}
    % \label{fig:OCSAL}
    % \end{figure*}

\textbf{Control Layer.} 
Cybersecurity/software security frameworks often define a set of controls(security requirements) intended to reduce the risk to a system. Framework authors typically organize these controls into a \textit{catalog}. The \textit{catalog model} is the basis for all other OSCAL models. Controls used in any other OSCAL model must first be defined in this model. 
Then, organizations and system owners identify which controls are applicable to a system/use case, which may include controls from more than one catalog.  
The \textit{profile model} for selecting, organizing, and tailoring a specific set of controls.
    A profile enables controls to be selected and tailored to express a baseline of controls. 
    
\textbf{Implementation Layer.} The OSCAL Implementation Layer focuses on the implementation of a system under a specific baseline as well as the individual components that may be incorporated into a system. 

\textbf{Assessment Layer.} 
The OSCAL Assessment Layer focuses on assessment activities, on communicating all assessment findings including supporting evidence, and identifying and managing the remediation of identified risks to a system identified as a result of assessment activities.
    
The Australian Signals Directorate (ASD) provides the Information Security Manual (ISM) in the OSCAL format\footnote{ISM OSCAL releases. \url{https://www.cyber.gov.au/resources-business-and-government/essential-cyber-security/ism/oscal}}. Additionally, NIST has published various OSCAL learning resources\footnote{OSCAL learning resources. \url{https://pages.nist.gov/OSCAL/learn/}} to help organizations understand the underlying concepts and practical applications of OSCAL.

\subsubsection{OSCAL-aligned Format for Software Security Operations}

%\sunny{clarify we focus solely on catalog model, and don't adapt other layers.}

%\sunny{may need to mention how we used OSCAL, fully or there are some modifications for this project....? It would be good to include a figure (e.g., OSCAL original - our models)}

\begin{figure*} 
    \centering
    \includegraphics[width=\textwidth]{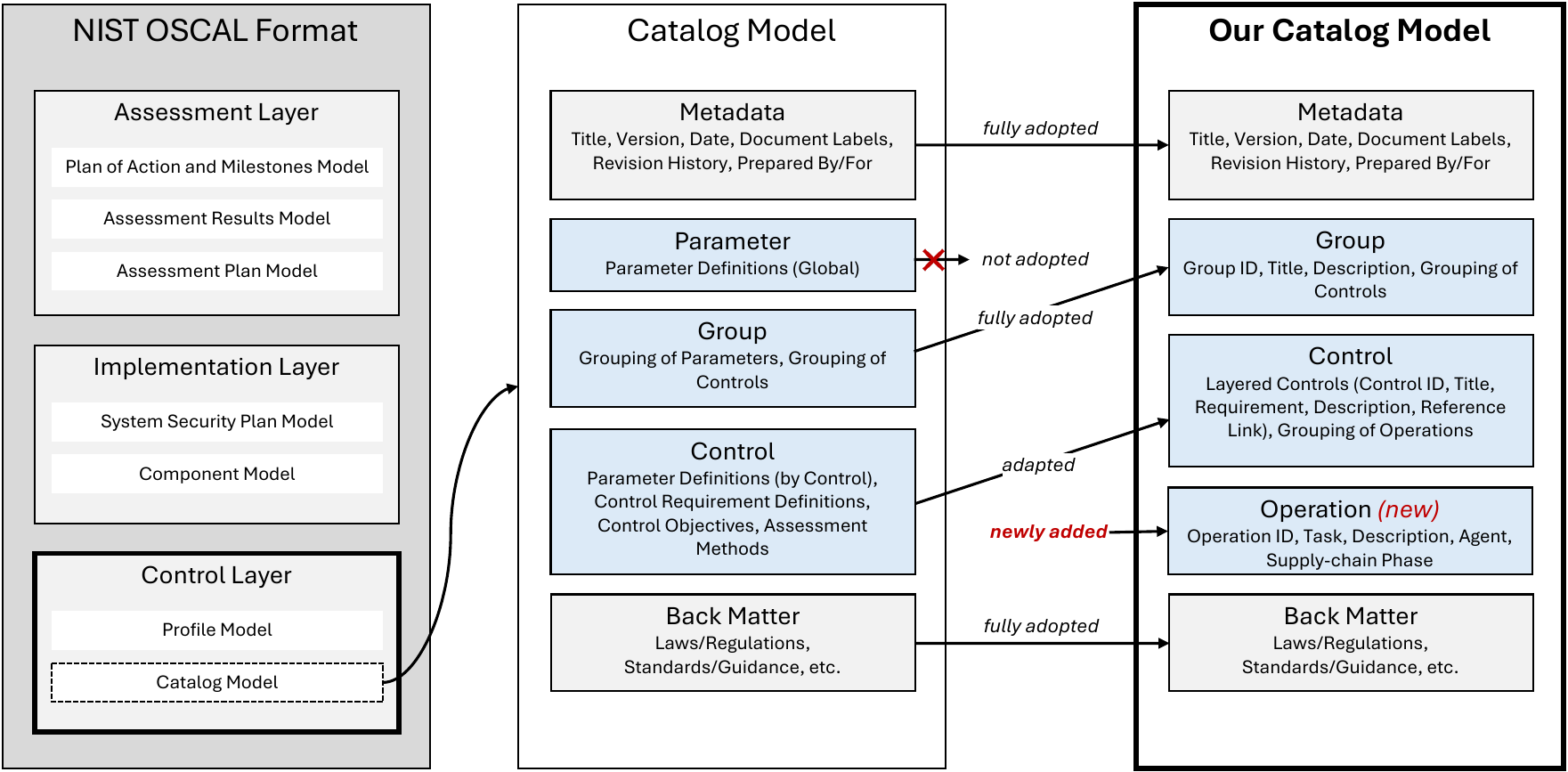}
    \caption{NIST OSCAL format (layers and models) and our catalog model.}
    \label{fig:catalog}
\end{figure*}

%In this paper, we present the OSCAL control layer in a machine-readable format, while the implementation and assessment layers will be introduced upon the completion of our evaluation.
In this study, we adopted the \textit{control layer}, while excluding other layers, as implementation and assessment are beyond the scope of this project.
In particular, we use the \textit{Catalog model} in the Control layer to represent our holistic mapping framework. Figure \ref{fig:catalog} provides an overview of the structure of our catalog model illustrating how the NIST OSCAL model was applied in this study. The primary objective of using OSCAL catalogs is to define organized sets of security controls. Therefore, the OSCAL catalog model offers the ability to group related controls, and the ability to define individual controls as well. 
To transform our mapping into the OSCAL catalog model, we introduced a newly defined operation component, which is nested within individual controls.
As a result, as highlighted in Figure \ref{fig:catalog}, our modified catalog model follows a top-down structure, from a top-level group to a set of security controls and their associated operations.

% \begin{figure*} [htb]
%     \centering
%     \includegraphics[width=0.3\textwidth]{Fig/Catalog.drawio.pdf}
%     \caption{Catalog Model Overview. \sunny{Can we merge Figure 10 and 11, side-by-side, using sub-figure?}}
%     \label{fig:Catalog Model}
% \end{figure*}

Representing the mapping structure in a machine-readable format requires adding structure around textual information, enabling each data point to be easily identified and processed by a machine.
In the following, we provide a simple OSCAL catalog model example for the "Secure Software Environment" goal. 

\begin{figure*} [htb]
    \centering
    \includegraphics[width=\textwidth]{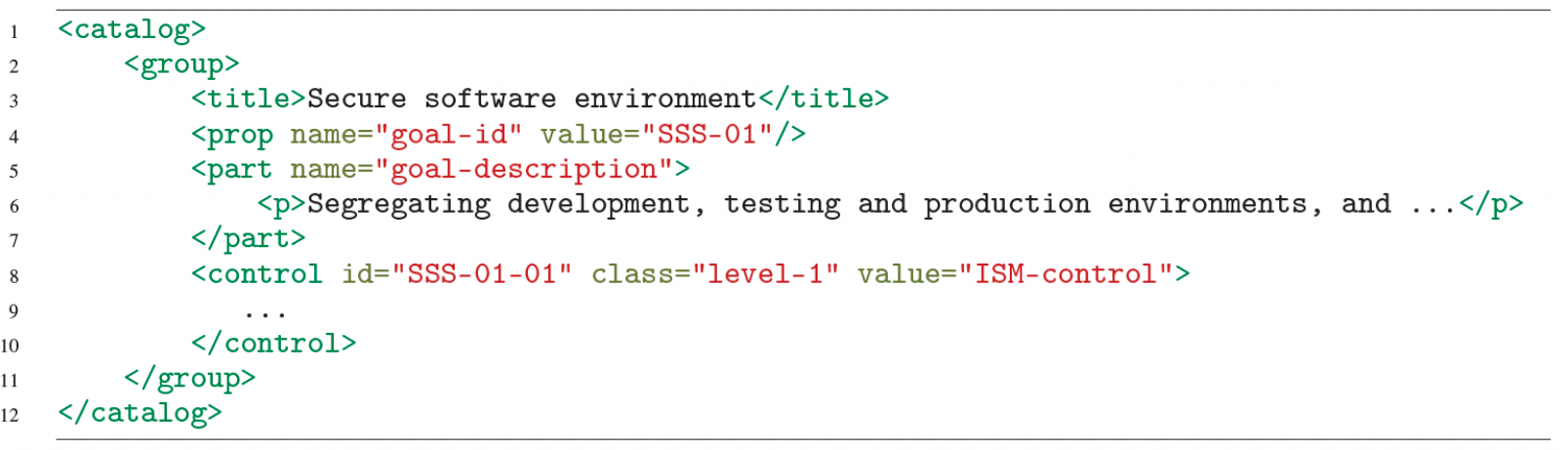}
    \caption{A top-level strategic goal, in OSCAL.}
    \label{fig:group}
\end{figure*}

\textbf{A top-level strategic goal.} 
The ISM guidelines for software development define four top-level strategic sub-goals which represented as \emph{<group>} in OSCAL format. Each \emph{<group>} (A snippet for a single group see Figure \ref{fig:group}) includes:
a unique identifier (e.g., goal-id: "SSS-01"), a title (e.g., "Secure Software Environment"), a goal description, and a set of nested security \emph{<control>} which represented as different levels security requirements.

\begin{figure*} [htb]
    \centering
    \includegraphics[width=\textwidth]{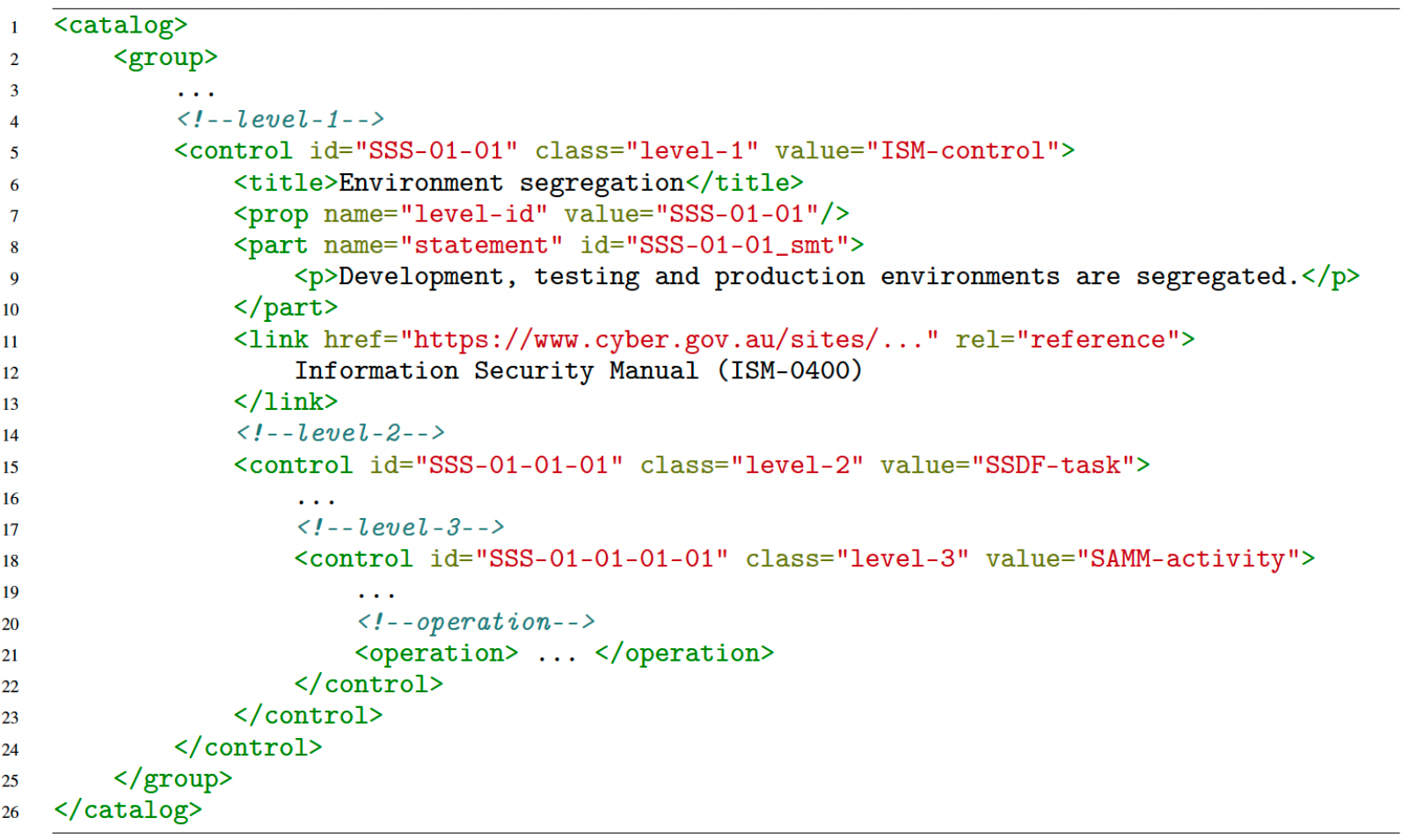}
    \caption{Three level security controls (requirements), in OSCAL}
    \label{fig:controls}
\end{figure*}

\textbf{Three level security requirements.}
Level-1 security requirements are represented as a single \emph{<control>}, with their subcontrols (Level-2 and Level-3 security requirements) nested in OSCAL format. These elements are structured under the sub-goal "Secure Software Environment", ensuring a standardized and machine-readable representation.
Each \emph{<control>} (see Figure \ref{fig:controls}) encompasses multiple nested Level-1 to Level-3 requirements. It includes: a title (e.g., \emph{<title>}Environment segregation\emph{</title>}), an identifier (ID) (e.g., "SSS-01-01"), the original security requirement statement, an interpreted description, reference link resources (e.g., \emph{<link>})  and generated several step operations (although not in this example).

\begin{figure*}
    \centering
    \includegraphics[width=\textwidth]{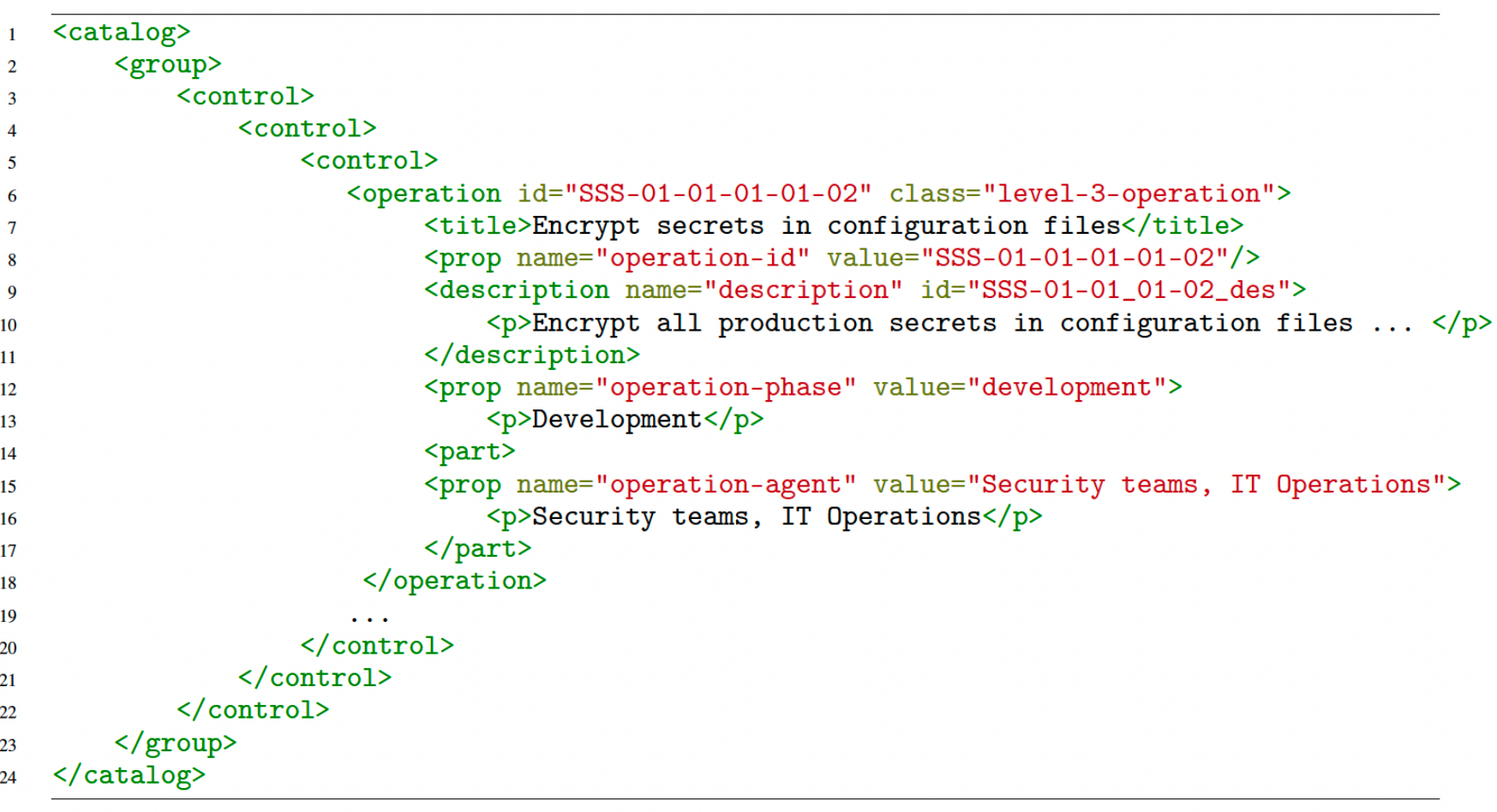}
    \caption{Generated Operations, in OSCAL}
    \label{fig:operations}
\end{figure*}

\textbf{Generated Operations.}
In the OSCAL catalog model, both group and control are redefined. To incorporate the Operations from our proposed Reference Architecture into the catalog model, we introduce a newly defined element within the security control: \emph{<operation>}.
The \emph{<operation>} (see Figure \ref{fig:operations}) including an operation identifier (e.g., id="SSS-01-01-01-01-02"), a title ("Ensure Secret Isolation "), the description, and alongside the responsible parties ("operation-agent") and the timeline ("operation-phase") in which they are conducted. 

\begin{figure*}
    \centering
    \includegraphics[width=\textwidth]{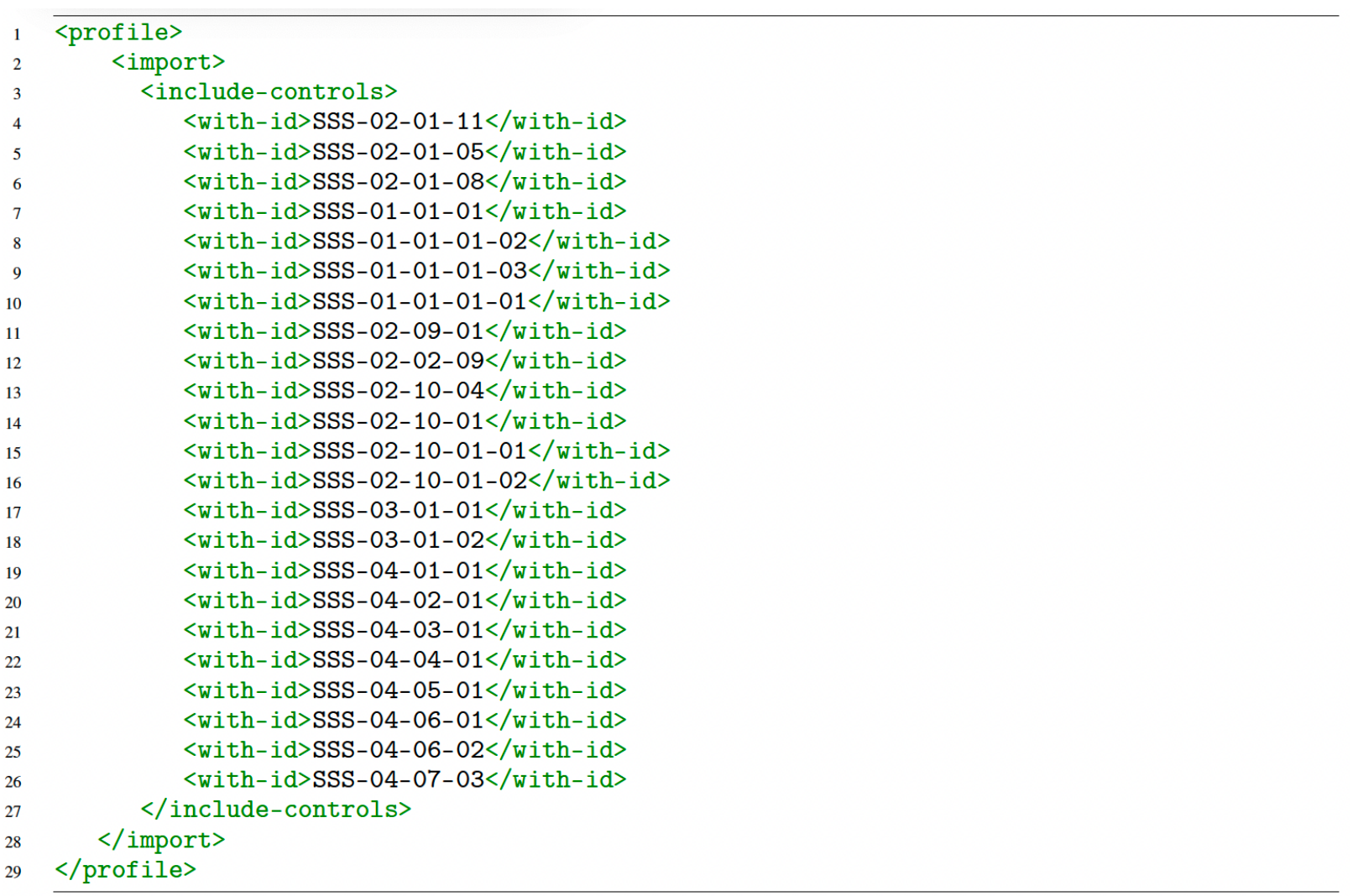}
    \caption{Profile Model (Log4j Security Checklist), in OSCAL}
    \label{fig:profile}
\end{figure*}

While the \textit{Profile model} is not the primary focus of this study, we briefly illustrate how our catalog model can be used to address specific security concerns, such as \textit{Log4j incident}.
In the previous section, we presented Table \ref{tab:checklist} including a recommended security checklist, consisting of 23 requirements, generated using our mapping framework.
An example of the OSCAL profile model format for the Log4j security checklist is provided in Figure \ref{fig:profile}.

\subsection{Theoretical Implications}

\textbf{Goal-driven Operationalization of Security Requirements.}
This study advances the theory of software security requirements by demonstrating how high-level goals can be systematically decomposed into structured, actionable components. Using a goal-oriented requirements engineering (GORE) approach, specifically the KAOS model, we offer a multi-layered mapping framework that links strategic intent to operational tasks. This structure provides semantic clarity, traceability, and contextual relevance, addressing a common limitation in existing security frameworks that often lack implementation-level detail.

\textbf{Synthesis of Fragmented Security Knowledge.}
This research synthesizes fragmented and often ambiguous security guidance—spread across multiple industry and government frameworks—into a unified, coherent model. By integrating inputs from key security frameworks (e.g., ISM, SSDF, SAMM, SLSA), our framework not only supports alignment across differing standards but also fills conceptual and operational gaps between them. This contributes to theory building by offering a more holistic view of software security as a multi-framework, life-cycle-spanning concern.

\textbf{Bridging Abstract Frameworks and Real-world Practices.}
While most security frameworks operate at a high level of abstraction, our framework introduces a middle layer that connects these abstract requirements to practical, real-world operations. By defining over 400 operations mapped to specific agents and software supply chain phases, we provide a tangible model for how organizations can implement and assess security requirements across the software life-cycle. This operational granularity contributes to the theoretical understanding of how security practices are distributed and contextualized across organizational roles and development stages.

\textbf{A Foundation for Machine-readable Security Modeling.}
The integration of our mapping framework into an extended, OSCAL-based machine-readable format opens new theoretical directions for automation, compliance, and interoperability. By introducing a new \textit{<operation>} component and aligning with existing OSCAL models, we demonstrate how structured requirement models can evolve into dynamic, tool-friendly artifacts. This supports research on formalizing and automating security controls, and offers a foundation for future work in standards-based, life-cycle-aware security assurance.

\section{Conclusion} \label{sec:Conclusion}
%\sunny{This section needs to be rewritten for journal submission.}
This paper introduces a holistic software security mapping framework that operationalizes security requirements across
three levels, linking regulatory guidelines to technical implementation tasks. Through collaborative research and goal-oriented modeling, we identify four foundational security goals and map them to refined requirements and operational actions. 
Our approach improves traceability, responsibility assignment, and practical implementation of secure software practices. 
Supported by a web-based navigation tool and real-world case checklist, this framework offers a robust foundation for organizations aiming to enhance software supply chain security. Future work will focus on evaluation, community feedback, and extending machine-readable formats for broader adoption.

%Bibliography
\bibliographystyle{unsrt}  
\bibliography{mapping}

\end{document}